\documentclass[twocolumn]{aastex631}

\usepackage{booktabs}
\usepackage{multirow}
\usepackage{float}
\usepackage{amsmath}
\usepackage{cleveref}
\usepackage{hyperref}

\usepackage{enumitem}
\usepackage{booktabs}

\allowdisplaybreaks

\begin{document}

\title{Detection of Millimeter-Wavelength Flares from Two Accreting White Dwarf Systems in the SPT-3G Galactic Plane Survey}

\author[0009-0007-0072-0445]{Y.~Wan}
\affiliation{Department of Astronomy, University of Illinois Urbana-Champaign, 1002 West Green Street, Urbana, IL, 61801, USA}
\affiliation{Center for AstroPhysical Surveys, National Center for Supercomputing Applications, Urbana, IL, 61801, USA}

\author[0000-0001-7192-3871]{J.~D.~Vieira}
\affiliation{Department of Astronomy, University of Illinois Urbana-Champaign, 1002 West Green Street, Urbana, IL, 61801, USA}
\affiliation{Department of Physics, University of Illinois Urbana-Champaign, 1110 West Green Street, Urbana, IL, 61801, USA}
\affiliation{Center for AstroPhysical Surveys, National Center for Supercomputing Applications, Urbana, IL, 61801, USA}

\author[0000-0002-5397-9035]{P.~M.~Chichura}
\affiliation{Department of Physics, University of Chicago, 5640 South Ellis Avenue, Chicago, IL, 60637, USA}
\affiliation{Kavli Institute for Cosmological Physics, University of Chicago, 5640 South Ellis Avenue, Chicago, IL, 60637, USA}

\author{T.~J.~Maccarone}
\affiliation{Department of Physics \& Astronomy, Texas Tech University, Box 41051, Lubbock, TX, 79409, USA}

\author[0000-0002-4435-4623]{A.~J.~Anderson}
\affiliation{Fermi National Accelerator Laboratory, MS209, P.O. Box 500, Batavia, IL, 60510, USA}
\affiliation{Kavli Institute for Cosmological Physics, University of Chicago, 5640 South Ellis Avenue, Chicago, IL, 60637, USA}
\affiliation{Department of Astronomy and Astrophysics, University of Chicago, 5640 South Ellis Avenue, Chicago, IL, 60637, USA}

\author{B.~Ansarinejad}
\affiliation{School of Physics, University of Melbourne, Parkville, VIC 3010, Australia}

\author[0000-0002-8935-9882]{A.~Anumarlapudi}
\affiliation{Department of Physics and Astronomy, University of North Carolina at Chapel Hill, Chapel Hill, NC 27599-3255, USA}

\author[0000-0002-0517-9842]{M.~Archipley}
\affiliation{Department of Astronomy and Astrophysics, University of Chicago, 5640 South Ellis Avenue, Chicago, IL, 60637, USA}
\affiliation{Kavli Institute for Cosmological Physics, University of Chicago, 5640 South Ellis Avenue, Chicago, IL, 60637, USA}

\author[0000-0001-6899-1873]{L.~Balkenhol}
\affiliation{Sorbonne Universit\'e, CNRS, UMR 7095, Institut d'Astrophysique de Paris, 98 bis bd Arago, 75014 Paris, France}

\author[0000-0001-9103-9354]{P.~S.~Barry}
\affiliation{School of Physics and Astronomy, Cardiff University, Cardiff, CF24 3AA, UK}

\author{K.~Benabed}
\affiliation{Sorbonne Universit\'e, CNRS, UMR 7095, Institut d'Astrophysique de Paris, 98 bis bd Arago, 75014 Paris, France}

\author[0000-0001-5868-0748]{A.~N.~Bender}
\affiliation{High-Energy Physics Division, Argonne National Laboratory, 9700 South Cass Avenue, Lemont, IL, 60439, USA}
\affiliation{Kavli Institute for Cosmological Physics, University of Chicago, 5640 South Ellis Avenue, Chicago, IL, 60637, USA}
\affiliation{Department of Astronomy and Astrophysics, University of Chicago, 5640 South Ellis Avenue, Chicago, IL, 60637, USA}

\author[0000-0002-5108-6823]{B.~A.~Benson}
\affiliation{Fermi National Accelerator Laboratory, MS209, P.O. Box 500, Batavia, IL, 60510, USA}
\affiliation{Kavli Institute for Cosmological Physics, University of Chicago, 5640 South Ellis Avenue, Chicago, IL, 60637, USA}
\affiliation{Department of Astronomy and Astrophysics, University of Chicago, 5640 South Ellis Avenue, Chicago, IL, 60637, USA}

\author[0000-0003-4847-3483]{F.~Bianchini}
\affiliation{Kavli Institute for Particle Astrophysics and Cosmology, Stanford University, 452 Lomita Mall, Stanford, CA, 94305, USA}
\affiliation{Department of Physics, Stanford University, 382 Via Pueblo Mall, Stanford, CA, 94305, USA}
\affiliation{SLAC National Accelerator Laboratory, 2575 Sand Hill Road, Menlo Park, CA, 94025, USA}

\author[0000-0001-7665-5079]{L.~E.~Bleem}
\affiliation{High-Energy Physics Division, Argonne National Laboratory, 9700 South Cass Avenue, Lemont, IL, 60439, USA}
\affiliation{Kavli Institute for Cosmological Physics, University of Chicago, 5640 South Ellis Avenue, Chicago, IL, 60637, USA}
\affiliation{Department of Astronomy and Astrophysics, University of Chicago, 5640 South Ellis Avenue, Chicago, IL, 60637, USA}

\author[0000-0002-8051-2924]{F.~R.~Bouchet}
\affiliation{Sorbonne Universit\'e, CNRS, UMR 7095, Institut d'Astrophysique de Paris, 98 bis bd Arago, 75014 Paris, France}

\author{L.~Bryant}
\affiliation{Enrico Fermi Institute, University of Chicago, 5640 South Ellis Avenue, Chicago, IL, 60637, USA}

\author[0000-0003-3483-8461]{E.~Camphuis}
\affiliation{Sorbonne Universit\'e, CNRS, UMR 7095, Institut d'Astrophysique de Paris, 98 bis bd Arago, 75014 Paris, France}

\author{M.~G.~Campitiello}
\affiliation{High-Energy Physics Division, Argonne National Laboratory, 9700 South Cass Avenue, Lemont, IL, 60439, USA}

\author[0000-0002-2044-7665]{J.~E.~Carlstrom}
\affiliation{Kavli Institute for Cosmological Physics, University of Chicago, 5640 South Ellis Avenue, Chicago, IL, 60637, USA}
\affiliation{Enrico Fermi Institute, University of Chicago, 5640 South Ellis Avenue, Chicago, IL, 60637, USA}
\affiliation{Department of Physics, University of Chicago, 5640 South Ellis Avenue, Chicago, IL, 60637, USA}
\affiliation{High-Energy Physics Division, Argonne National Laboratory, 9700 South Cass Avenue, Lemont, IL, 60439, USA}
\affiliation{Department of Astronomy and Astrophysics, University of Chicago, 5640 South Ellis Avenue, Chicago, IL, 60637, USA}

\author{C.~L.~Chang}
\affiliation{High-Energy Physics Division, Argonne National Laboratory, 9700 South Cass Avenue, Lemont, IL, 60439, USA}
\affiliation{Kavli Institute for Cosmological Physics, University of Chicago, 5640 South Ellis Avenue, Chicago, IL, 60637, USA}
\affiliation{Department of Astronomy and Astrophysics, University of Chicago, 5640 South Ellis Avenue, Chicago, IL, 60637, USA}

\author{P.~Chaubal}
\affiliation{School of Physics, University of Melbourne, Parkville, VIC 3010, Australia}

\author{A.~Chokshi}
\affiliation{University of Chicago, 5640 South Ellis Avenue, Chicago, IL, 60637, USA}

\author[0000-0002-3091-8790]{T.-L.~Chou}
\affiliation{Department of Astronomy and Astrophysics, University of Chicago, 5640 South Ellis Avenue, Chicago, IL, 60637, USA}
\affiliation{Kavli Institute for Cosmological Physics, University of Chicago, 5640 South Ellis Avenue, Chicago, IL, 60637, USA}
\affiliation{National Taiwan University, No. 1, Sec. 4, Roosevelt Road, Taipei 106319, Taiwan}

\author{A.~Coerver}
\affiliation{Department of Physics, University of California, Berkeley, CA, 94720, USA}

\author[0000-0001-9000-5013]{T.~M.~Crawford}
\affiliation{Department of Astronomy and Astrophysics, University of Chicago, 5640 South Ellis Avenue, Chicago, IL, 60637, USA}
\affiliation{Kavli Institute for Cosmological Physics, University of Chicago, 5640 South Ellis Avenue, Chicago, IL, 60637, USA}

\author[0000-0002-3760-2086]{C.~Daley}
\affiliation{Universit\'e Paris-Saclay, Universit\'e Paris Cit\'e, CEA, CNRS, AIM, 91191, Gif-sur-Yvette, France}
\affiliation{Department of Astronomy, University of Illinois Urbana-Champaign, 1002 West Green Street, Urbana, IL, 61801, USA}

\author{T.~de~Haan}
\affiliation{High Energy Accelerator Research Organization (KEK), Tsukuba, Ibaraki 305-0801, Japan}

\author{K.~R.~Dibert}
\affiliation{Department of Astronomy and Astrophysics, University of Chicago, 5640 South Ellis Avenue, Chicago, IL, 60637, USA}
\affiliation{Kavli Institute for Cosmological Physics, University of Chicago, 5640 South Ellis Avenue, Chicago, IL, 60637, USA}

\author{M.~A.~Dobbs}
\affiliation{Department of Physics and McGill Space Institute, McGill University, 3600 Rue University, Montreal, Quebec H3A 2T8, Canada}
\affiliation{Canadian Institute for Advanced Research, CIFAR Program in Gravity and the Extreme Universe, Toronto, ON, M5G 1Z8, Canada}

\author{M.~Doohan}
\affiliation{School of Physics, University of Melbourne, Parkville, VIC 3010, Australia}

\author{A.~Doussot}
\affiliation{Sorbonne Universit\'e, CNRS, UMR 7095, Institut d'Astrophysique de Paris, 98 bis bd Arago, 75014 Paris, France}

\author[0000-0002-9962-2058]{D.~Dutcher}
\affiliation{Joseph Henry Laboratories of Physics, Jadwin Hall, Princeton University, Princeton, NJ 08544, USA}

\author{W.~Everett}
\affiliation{Department of Astrophysical and Planetary Sciences, University of Colorado, Boulder, CO, 80309, USA}

\author{C.~Feng}
\affiliation{Department of Physics, University of Illinois Urbana-Champaign, 1110 West Green Street, Urbana, IL, 61801, USA}

\author[0000-0002-4928-8813]{K.~R.~Ferguson}
\affiliation{Department of Physics and Astronomy, University of California, Los Angeles, CA, 90095, USA}
\affiliation{Department of Physics and Astronomy, Michigan State University, East Lansing, MI 48824, USA}

\author{K.~Fichman}
\affiliation{Department of Physics, University of Chicago, 5640 South Ellis Avenue, Chicago, IL, 60637, USA}
\affiliation{Kavli Institute for Cosmological Physics, University of Chicago, 5640 South Ellis Avenue, Chicago, IL, 60637, USA}

\author[0000-0002-7145-1824]{A.~Foster}
\affiliation{Joseph Henry Laboratories of Physics, Jadwin Hall, Princeton University, Princeton, NJ 08544, USA}

\author{S.~Galli}
\affiliation{Sorbonne Universit\'e, CNRS, UMR 7095, Institut d'Astrophysique de Paris, 98 bis bd Arago, 75014 Paris, France}

\author{A.~E.~Gambrel}
\affiliation{Kavli Institute for Cosmological Physics, University of Chicago, 5640 South Ellis Avenue, Chicago, IL, 60637, USA}

\author{R.~W.~Gardner}
\affiliation{Enrico Fermi Institute, University of Chicago, 5640 South Ellis Avenue, Chicago, IL, 60637, USA}

\author{F.~Ge}
\affiliation{Kavli Institute for Particle Astrophysics and Cosmology, Stanford University, 452 Lomita Mall, Stanford, CA, 94305, USA}
\affiliation{Department of Physics, Stanford University, 382 Via Pueblo Mall, Stanford, CA, 94305, USA}
\affiliation{Department of Physics \& Astronomy, University of California, One Shields Avenue, Davis, CA 95616, USA}

\author{N.~Goeckner-Wald}
\affiliation{Department of Physics, Stanford University, 382 Via Pueblo Mall, Stanford, CA, 94305, USA}
\affiliation{Kavli Institute for Particle Astrophysics and Cosmology, Stanford University, 452 Lomita Mall, Stanford, CA, 94305, USA}

\author[0000-0003-4245-2315]{R.~Gualtieri}
\affiliation{High-Energy Physics Division, Argonne National Laboratory, 9700 South Cass Avenue, Lemont, IL, 60439, USA}
\affiliation{Department of Physics and Astronomy, Northwestern University, 633 Clark St, Evanston, IL, 60208, USA}

\author[0000-0001-7593-3962]{F.~Guidi}
\affiliation{Sorbonne Universit\'e, CNRS, UMR 7095, Institut d'Astrophysique de Paris, 98 bis bd Arago, 75014 Paris, France}

\author{S.~Guns}
\affiliation{Department of Physics, University of California, Berkeley, CA, 94720, USA}

\author{N.~W.~Halverson}
\affiliation{CASA, Department of Astrophysical and Planetary Sciences, University of Colorado, Boulder, CO, 80309, USA }
\affiliation{Department of Physics, University of Colorado, Boulder, CO, 80309, USA}

\author[0000-0003-1880-2733]{E.~Hivon}
\affiliation{Sorbonne Universit\'e, CNRS, UMR 7095, Institut d'Astrophysique de Paris, 98 bis bd Arago, 75014 Paris, France}

\author[0000-0002-0463-6394]{G.~P.~Holder}
\affiliation{Department of Physics, University of Illinois Urbana-Champaign, 1110 West Green Street, Urbana, IL, 61801, USA}

\author{W.~L.~Holzapfel}
\affiliation{Department of Physics, University of California, Berkeley, CA, 94720, USA}

\author{J.~C.~Hood}
\affiliation{Kavli Institute for Cosmological Physics, University of Chicago, 5640 South Ellis Avenue, Chicago, IL, 60637, USA}

\author{A.~Hryciuk}
\affiliation{Department of Physics, University of Chicago, 5640 South Ellis Avenue, Chicago, IL, 60637, USA}
\affiliation{Kavli Institute for Cosmological Physics, University of Chicago, 5640 South Ellis Avenue, Chicago, IL, 60637, USA}

\author[0000-0003-3595-0359]{N.~Huang}
\affiliation{Department of Physics, University of California, Berkeley, CA, 94720, USA}

\author[0000-0001-6295-2881]{D.~L.~Kaplan}
\affiliation{Department of Physics, University of Wisconsin-Milwaukee, P.O. Box 413, Milwaukee, WI 53201, USA}

\author{F.~K\'eruzor\'e}
\affiliation{High-Energy Physics Division, Argonne National Laboratory, 9700 South Cass Avenue, Lemont, IL, 60439, USA}

\author[0000-0002-8388-4950]{A.~R.~Khalife}
\affiliation{Sorbonne Universit\'e, CNRS, UMR 7095, Institut d'Astrophysique de Paris, 98 bis bd Arago, 75014 Paris, France}

\author{L.~Knox}
\affiliation{Department of Physics \& Astronomy, University of California, One Shields Avenue, Davis, CA 95616, USA}

\author{M.~Korman}
\affiliation{Department of Physics, Case Western Reserve University, Cleveland, OH, 44106, USA}

\author{K.~Kornoelje}
\affiliation{Department of Astronomy and Astrophysics, University of Chicago, 5640 South Ellis Avenue, Chicago, IL, 60637, USA}
\affiliation{Kavli Institute for Cosmological Physics, University of Chicago, 5640 South Ellis Avenue, Chicago, IL, 60637, USA}
\affiliation{High-Energy Physics Division, Argonne National Laboratory, 9700 South Cass Avenue, Lemont, IL, 60439, USA}

\author{C.-L.~Kuo}
\affiliation{Kavli Institute for Particle Astrophysics and Cosmology, Stanford University, 452 Lomita Mall, Stanford, CA, 94305, USA}
\affiliation{Department of Physics, Stanford University, 382 Via Pueblo Mall, Stanford, CA, 94305, USA}
\affiliation{SLAC National Accelerator Laboratory, 2575 Sand Hill Road, Menlo Park, CA, 94025, USA}

\author{K.~Levy}
\affiliation{School of Physics, University of Melbourne, Parkville, VIC 3010, Australia}

\author[0000-0002-4747-4276]{A.~E.~Lowitz}
\affiliation{Kavli Institute for Cosmological Physics, University of Chicago, 5640 South Ellis Avenue, Chicago, IL, 60637, USA}

\author{C.~Lu}
\affiliation{Department of Physics, University of Illinois Urbana-Champaign, 1110 West Green Street, Urbana, IL, 61801, USA}

\author[0009-0004-3143-1708]{G.~P.~Lynch}
\affiliation{Department of Physics \& Astronomy, University of California, One Shields Avenue, Davis, CA 95616, USA}

\author{A.~Maniyar}
\affiliation{Kavli Institute for Particle Astrophysics and Cosmology, Stanford University, 452 Lomita Mall, Stanford, CA, 94305, USA}
\affiliation{Department of Physics, Stanford University, 382 Via Pueblo Mall, Stanford, CA, 94305, USA}
\affiliation{SLAC National Accelerator Laboratory, 2575 Sand Hill Road, Menlo Park, CA, 94025, USA}

\author{E.~S.~Martsen}
\affiliation{Department of Astronomy and Astrophysics, University of Chicago, 5640 South Ellis Avenue, Chicago, IL, 60637, USA}
\affiliation{Kavli Institute for Cosmological Physics, University of Chicago, 5640 South Ellis Avenue, Chicago, IL, 60637, USA}

\author{F.~Menanteau}
\affiliation{Department of Astronomy, University of Illinois Urbana-Champaign, 1002 West Green Street, Urbana, IL, 61801, USA}
\affiliation{Center for AstroPhysical Surveys, National Center for Supercomputing Applications, Urbana, IL, 61801, USA}

\author[0000-0001-7317-0551]{M.~Millea}
\affiliation{Department of Physics, University of California, Berkeley, CA, 94720, USA}

\author{J.~Montgomery}
\affiliation{Department of Physics and McGill Space Institute, McGill University, 3600 Rue University, Montreal, Quebec H3A 2T8, Canada}

\author{Y.~Nakato}
\affiliation{Department of Physics, Stanford University, 382 Via Pueblo Mall, Stanford, CA, 94305, USA}

\author{T.~Natoli}
\affiliation{Kavli Institute for Cosmological Physics, University of Chicago, 5640 South Ellis Avenue, Chicago, IL, 60637, USA}

\author[0000-0002-5254-243X]{G.~I.~Noble}
\affiliation{Dunlap Institute for Astronomy \& Astrophysics, University of Toronto, 50 St. George Street, Toronto, ON, M5S 3H4, Canada}
\affiliation{David A. Dunlap Department of Astronomy \& Astrophysics, University of Toronto, 50 St. George Street, Toronto, ON, M5S 3H4, Canada}

\author{Y.~Omori}
\affiliation{Department of Astronomy and Astrophysics, University of Chicago, 5640 South Ellis Avenue, Chicago, IL, 60637, USA}
\affiliation{Kavli Institute for Cosmological Physics, University of Chicago, 5640 South Ellis Avenue, Chicago, IL, 60637, USA}

\author{A.~Ouellette}
\affiliation{Department of Physics, University of Illinois Urbana-Champaign, 1110 West Green Street, Urbana, IL, 61801, USA}

\author[0000-0002-6164-9861]{Z.~Pan}
\affiliation{High-Energy Physics Division, Argonne National Laboratory, 9700 South Cass Avenue, Lemont, IL, 60439, USA}
\affiliation{Kavli Institute for Cosmological Physics, University of Chicago, 5640 South Ellis Avenue, Chicago, IL, 60637, USA}
\affiliation{Department of Physics, University of Chicago, 5640 South Ellis Avenue, Chicago, IL, 60637, USA}

\author{P.~Paschos}
\affiliation{Enrico Fermi Institute, University of Chicago, 5640 South Ellis Avenue, Chicago, IL, 60637, USA}

\author[0000-0001-7946-557X]{K.~A.~Phadke}
\affiliation{Department of Astronomy, University of Illinois Urbana-Champaign, 1002 West Green Street, Urbana, IL, 61801, USA}
\affiliation{Center for AstroPhysical Surveys, National Center for Supercomputing Applications, Urbana, IL, 61801, USA}
\affiliation{NSF-Simons AI Institute for the Sky (SkAI), 172 E. Chestnut St., Chicago, IL 60611, USA}

\author{A.~W.~Pollak}
\affiliation{University of Chicago, 5640 South Ellis Avenue, Chicago, IL, 60637, USA}

\author{K.~Prabhu}
\affiliation{Department of Physics \& Astronomy, University of California, One Shields Avenue, Davis, CA 95616, USA}

\author{W.~Quan}
\affiliation{High-Energy Physics Division, Argonne National Laboratory, 9700 South Cass Avenue, Lemont, IL, 60439, USA}
\affiliation{Department of Physics, University of Chicago, 5640 South Ellis Avenue, Chicago, IL, 60637, USA}
\affiliation{Kavli Institute for Cosmological Physics, University of Chicago, 5640 South Ellis Avenue, Chicago, IL, 60637, USA}

\author{M.~Rahimi}
\affiliation{School of Physics, University of Melbourne, Parkville, VIC 3010, Australia}

\author[0000-0003-3953-1776]{A.~Rahlin}
\affiliation{Department of Astronomy and Astrophysics, University of Chicago, 5640 South Ellis Avenue, Chicago, IL, 60637, USA}
\affiliation{Kavli Institute for Cosmological Physics, University of Chicago, 5640 South Ellis Avenue, Chicago, IL, 60637, USA}

\author[0000-0003-2226-9169]{C.~L.~Reichardt}
\affiliation{School of Physics, University of Melbourne, Parkville, VIC 3010, Australia}

\author{M.~Rouble}
\affiliation{Department of Physics and McGill Space Institute, McGill University, 3600 Rue University, Montreal, Quebec H3A 2T8, Canada}

\author{J.~E.~Ruhl}
\affiliation{Department of Physics, Case Western Reserve University, Cleveland, OH, 44106, USA}

\author{E.~Schiappucci}
\affiliation{School of Physics, University of Melbourne, Parkville, VIC 3010, Australia}

\author{A.~Simpson}
\affiliation{Department of Astronomy and Astrophysics, University of Chicago, 5640 South Ellis Avenue, Chicago, IL, 60637, USA}
\affiliation{Kavli Institute for Cosmological Physics, University of Chicago, 5640 South Ellis Avenue, Chicago, IL, 60637, USA}

\author[0000-0001-6155-5315]{J.~A.~Sobrin}
\affiliation{Fermi National Accelerator Laboratory, MS209, P.O. Box 500, Batavia, IL, 60510, USA}
\affiliation{Kavli Institute for Cosmological Physics, University of Chicago, 5640 South Ellis Avenue, Chicago, IL, 60637, USA}

\author{A.~A.~Stark}
\affiliation{Center for Astrophysics \textbar{} Harvard \& Smithsonian, 60 Garden Street, Cambridge, MA, 02138, USA}

\author{J.~Stephen}
\affiliation{Enrico Fermi Institute, University of Chicago, 5640 South Ellis Avenue, Chicago, IL, 60637, USA}

\author[0000-0002-2077-6004]{C.~Tandoi}
\affiliation{Department of Astronomy, University of Illinois Urbana-Champaign, 1002 West Green Street, Urbana, IL, 61801, USA}

\author{B.~Thorne}
\affiliation{Department of Physics \& Astronomy, University of California, One Shields Avenue, Davis, CA 95616, USA}

\author{C.~Trendafilova}
\affiliation{Center for AstroPhysical Surveys, National Center for Supercomputing Applications, Urbana, IL, 61801, USA}

\author[0000-0002-6805-6188]{C.~Umilta}
\affiliation{Department of Physics, University of Illinois Urbana-Champaign, 1110 West Green Street, Urbana, IL, 61801, USA}

\author[0009-0009-3168-092X]{A.~Vitrier}
\affiliation{Sorbonne Universit\'e, CNRS, UMR 7095, Institut d'Astrophysique de Paris, 98 bis bd Arago, 75014 Paris, France}

\author[0000-0002-3157-0407]{N.~Whitehorn}
\affiliation{Department of Physics and Astronomy, Michigan State University, East Lansing, MI 48824, USA}

\author[0000-0001-5411-6920]{W.~L.~K.~Wu}
\affiliation{Kavli Institute for Particle Astrophysics and Cosmology, Stanford University, 452 Lomita Mall, Stanford, CA, 94305, USA}
\affiliation{SLAC National Accelerator Laboratory, 2575 Sand Hill Road, Menlo Park, CA, 94025, USA}

\author{M.~R.~Young}
\affiliation{Fermi National Accelerator Laboratory, MS209, P.O. Box 500, Batavia, IL, 60510, USA}
\affiliation{Kavli Institute for Cosmological Physics, University of Chicago, 5640 South Ellis Avenue, Chicago, IL, 60637, USA}

\author{J.~A.~Zebrowski}
\affiliation{Kavli Institute for Cosmological Physics, University of Chicago, 5640 South Ellis Avenue, Chicago, IL, 60637, USA}
\affiliation{Department of Astronomy and Astrophysics, University of Chicago, 5640 South Ellis Avenue, Chicago, IL, 60637, USA}
\affiliation{Fermi National Accelerator Laboratory, MS209, P.O. Box 500, Batavia, IL, 60510, USA}



\begin{abstract}

Blind discoveries of millimeter-wave (mm-wave) transient events in non-targeted surveys, as opposed to follow-up or pointed observations, have only become possible in the past decade using cosmic microwave background surveys. Here we present the first results from the SPT-3G Galactic Plane Survey---the first dedicated high-sensitivity, wide-field, time-domain, mm-wave survey of the Galactic Plane, conducted with the South Pole Telescope (SPT) using the SPT-3G camera. The survey field covers approximately 100~$\text{deg}^2$ near the Galactic center. In 2023 and 2024, this survey consists of roughly 1,500 individual 20-minute observations in three bands centered at 95, 150, and 220~GHz, with plans for more observations in the coming years. We report the detection of two transient events exceeding a 5$\sigma$ threshold in both the 95 and 150~GHz bands in the first two years of SPT-3G Galactic Plane Survey data. Both events are unpolarized and exhibit durations of approximately one day, with peak flux densities at 150~GHz of at least 50~mJy. The peak isotropic luminosities at 150~GHz are on the order of $10^{31}~\text{erg}~\text{s}^{-1}$. Both events are associated with previously identified accreting white dwarfs. Magnetic reconnection in the accretion disk is a likely explanation for the observed millimeter flares. In the future, we plan to expand the transient search in the Galactic Plane by lowering the detection threshold, enabling single-band detections, analyzing lightcurves on a range of timescales, and including additional data from future observations.




\end{abstract}



\section{Introduction}

The millimeter-wave (mm-wave) transient sky contains information on a wide range of astrophysical objects and phenomena, including stellar flares, varying active galactic nuclei (AGN), long-duration gamma-ray bursts, accreting white dwarfs, and more \citep{Eftekhari_2022,Holder_2019}. The transient sky has traditionally been observed at optical, X-ray, and other wavelengths, while the mm-wave sky remained largely unexplored until a decade ago due to technical limitations, including limited detector sensitivity and angular resolution \citep{Metzger_2015}. Current and future cosmic microwave background (CMB) instruments, such as the South Pole Telescope (SPT; \citealt{Carlstrom_2011}), the Atacama Cosmology Telescope (ACT; \citealt{Swetz_2011}), and the Simons Observatory (SO; \citealt{Lee_2019}), offer sufficient sensitivity and resolution to open a new window into the time-variable mm-wave sky.

Surveys with the SPT, a 10-m diameter, wide-field, mm-wave survey telescope \citep{Carlstrom_2011,Holder_2019} have played a key role in exploring this new discovery space. Transient source detection outside of the Galactic Plane has already been performed with the SPT in recent years. The first systematic search was carried out using one year of data during the 2012-2013 observing season covering a 100~$\text{deg}^{2}$ patch of sky with SPTpol (the second-generation camera on the SPT), in which one transient candidate was discovered and hypothesized to be a gamma-ray burst afterglow \citep{Whitehorn_2016}. Another transient search was conducted using 2020 data covering a larger field of 1500~$\text{deg}^{2}$ with SPT-3G (the third-generation camera on the SPT), yielding the detection of 13 stellar flares from eight flaring stars and two extragalactic sources, which are likely AGN \citep{Guns_2021}. A flare star catalog was published using four years of data from 2019-2022, consisting of 111 stellar flares from 66 stars \citep{Tandoi_2024}. Other SPT work on variable and transient sources includes focused studies of AGN \citep{Hood_2023}, asteroids \citep{Chichura_2022}, and satellites \citep{Foster_2025}. To date, the SPT has reported over a hundred transient events, with the majority being Galactic stellar flares, along with a few extragalactic transients.




Similar results for transient detections have been reported by ACT. Three transient sources were serendipitously discovered during the search for Planet 9, all of which are associated with Galactic flaring stars \citep{Naess_2021_transients,Naess_2021_planet9}. A systematic search was conducted using three years of data with three-day maps, yielding the detection of 14 stellar flares at 11 unique locations \citep{Li_2023}. Another systematic search was performed on single observation maps from 2017 to 2022, resulting in the discovery of 34 transient events at 27 unique locations, with all but two transient sources being associated with Galactic stars \citep{Biermann_2024}. No other non-targeted mm-wave transient searches have been published to date.

All previous SPT surveys have avoided the plane of the Milky Way Galaxy in order to minimize contamination from Galactic foregrounds. ACT observes a large area of the sky, including the Galaxy, but the Galactic Plane is masked during transient searches \citep{Biermann_2024}. There have been other sub/millimeter surveys towards the Galactic Plane, such as the \textit{Herschel} infrared Galactic Plane Survey (Hi-GAL; \citealt{Molinari_2016}), the APEX Telescope Large Area Survey of the Galaxy (ATLASGAL; \citealt{Schuller_2009}), the Bolocam Galactic Plane Survey (BGPS; \citealt{Aguirre_2011}), and the ACT Galactic center Survey \citep{Guan_2021}, but they were all static continuum surveys and did not include time-domain scientific analyses. The discovery of mostly Galactic transients in previous SPT transient searches \citep{Guns_2021, Tandoi_2024} and the lack of mm-wave time-domain exploration towards the Galactic Plane are the main motivations for the SPT-3G Galactic Plane Survey. 

There are classes of objects in the Galactic Plane known to emit strongly variable mm-wave emission, such as X-ray binaries \citep{Tetarenko}, classical novae \citep{Chomiuk}, magnetars \citep{2020ATel14001....1T}, and T Tauri stars \citep{Salter}, none of which are present in the high Galactic latitude fields observed to date. The SPT Galactic Plane time-domain survey has the potential to find mm-wave emission from such objects. Serendipitous discoveries of unexpected classes of transients are also possible, as the Galactic Plane remains largely unexplored in the time domain at millimeter wavelengths.

In this paper, we present the SPT-3G Galactic Plane Survey and its first results from this survey. We detected two transient events above a threshold of $5\sigma$ in both the 95 and 150~GHz bands simultaneously. Both events have timescales of approximately one day and we associate them with previously identified accreting white dwarf systems \citep{Shaw_2020, Munari_2021}. Accreting white dwarf systems, such as cataclysmic variables (CVs) and symbiotic stars, are binary systems in which a white dwarf accretes matter from a stellar companion. In the optical band, transient phenomena have been seen from these objects for centuries \citep{Warnerbook}, but their behavior at mm-wave is not well constrained, characterized, or understood. Mm-wave observations are crucial for building a complete picture of these objects by probing synchrotron emission from magnetic reconnection events within the systems \citep{Lazarian_2020}.

This paper is organized as follows: in \autoref{Sec:survey}, we describe the SPT and the SPT-3G Galactic Plane Survey; in \autoref{Sec:observations}, we outline the procedures for data processing and transient detection; in \autoref{Sec:results}, we present the results, including the lightcurves, polarization fractions, spectral indices, and source associations; in \autoref{Sec:discussion}, we discuss potential emission mechanisms for the detected transient events, compare our findings with previous searches, and provide an estimate of the event rates; and in \autoref{Sec:conclusion}, we conclude with a summary of the paper and final remarks.


\section{The Instrument and Survey}
\label{Sec:survey}



The SPT is a 10-meter, wide-field, mm-wave survey telescope located at the Amundsen-Scott South Pole Station in Antarctica \citep{Carlstrom_2011}. The SPT-3G receiver is the third-generation camera installed on the SPT and features roughly 16,000 bolometric detectors configured to simultaneously observe in three wavelength bands centered at 95, 150, and 220~GHz (corresponding to 3.2, 2.0, and 1.4~mm, respectively), with arcminute-scale resolution and polarization sensitivity \citep{Sobrin_2022}. SPT-3G has an instantaneous field of view of $1.9^\circ$ in diameter, which allows efficient coverage of large areas on the sky. 
The SPT is optimized to survey primary and secondary CMB anisotropies (e.g.,  \citealt{Camphuis_2025}). Moreover, the primary observing strategy employed with SPT-3G involves revisiting the same field regularly, making SPT-3G surveys powerful tool for time-domain science. 

The main SPT-3G cosmology survey covers an area of about 1500~$\text{deg}^2$ with declination (decl.) from $-42^{\circ}$ to $-70^{\circ}$ and right ascension (R.A.) from 20h40m0s to 3h20m0s \citep{Camphuis_2025}. This area has been observed with a near daily cadence during the majority of the SPT-3G observing period \citep{Dutcher_2021, Prabhu_2024}. Transient searches have been conducted on the main survey observations, resulting in the detection of dozens of flare stars and extragalactic sources \citep{Guns_2021, Tandoi_2024}. Most of the transient sources detected are nearby in the Galaxy (within 1,000~pc), which motivates the SPT-3G Galactic Plane Survey, as we can probe a different volume of the Milky Way with higher source density and different populations of sources. Main field observations are limited by the thickness of the Galactic Plane (a few hundred pc), whereas Galactic Plane observations can detect sources at larger distances, probing greater volumes for bright sources. Additionally, as there is an increase in stellar age with Galactic height \citep{Moreira_2025}, the volume probed by Galactic Plane observations likely contains different stellar populations. The SPT-3G Galactic Plane Survey is the first time the SPT has observed the Galactic Plane, as the other observing fields are chosen to avoid foreground contamination from the Galaxy, which would otherwise interfere with measurements of the CMB.



\begin{figure}
    \centering
    \includegraphics[width=9cm]{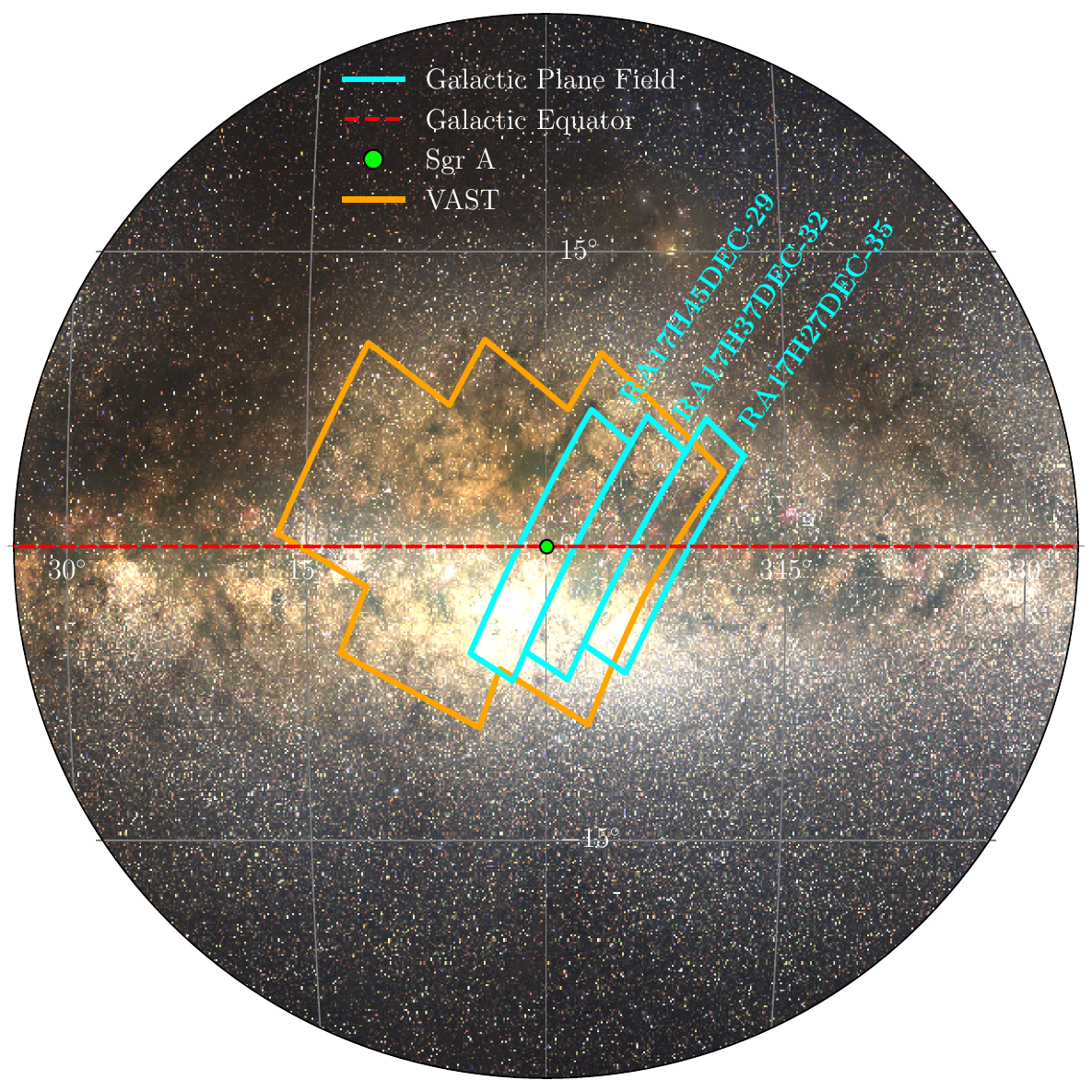}
  \caption{The footprint of the Galactic Plane field overlaid on Mellinger's all-sky panorama image of the Milky Way \citep{Mellinger_2009} in Galactic coordinates. The Galactic equator ($b = 0^\circ$) is marked by the dashed red line. The three SPT-3G Galactic subfields are shown by the blue boxes and the position of Sgr A is indicated by the green dot. For comparison, the orange region shows region 5 from the VAST Pilot Survey \citep{Murphy_2021}; see \autoref{sec:sourceassociation} for more details.}
  \label{Figure:footprint}
\end{figure}

The Galactic Plane field consists of three subfields---\textsc{ra17h45dec-29}, \textsc{ra17h37dec-32}, and \textsc{ra17h27dec-35}---as defined in \autoref{Table:nobs}. Sagittarius A (Sgr A), the radio source at the center of our Galaxy which contains the supermassive black hole Sagittarius A\textsuperscript{*} (Sgr A\textsuperscript{*}), is located at the center of the subfield \textsc{ra17h45dec-29}. Each subfield covers roughly 40~$\text{deg}^{2}$ and the uniformly covered region of the entire Galactic Plane field is about 100~$\text{deg}^{2}$. The highest declination (lowest elevation) the SPT can reach is constrained by atmospheric loading on the detectors, and the location of the field was initially designed to maximize the overlapping area with the Variables and Slow Transients Survey (VAST; \citealt{Murphy_2021, Wang_2022}) on the Australian Square Kilometre Array Pathfinder (ASKAP; \citealt{Hotan_2021}). \autoref{Figure:footprint} shows the location of these fields on the sky. 

When observing each subfield, the telescope starts at the lowest elevation of the subfield and scans back and forth at a constant speed along lines of constant elevation, which is equivalent to constant declination, as the telescope is located at the geographical South Pole. After finishing each scan, the telescope takes a step of 7.5$^{\prime}$ up in elevation and starts a new scan until covering the whole elevation range of the subfield. Each back and forth scan takes approximately 51 seconds and each observation consists of 24 scans on average. Thus, each observation takes approximately 20 minutes. A typical SPT-3G observing day is approximately 20.5 hours, including 16 hours of observing and 4.5 hours allocated to the fridge cycle required to cool the detectors to their operating temperature. In each Galactic Plane observing day, we observe all three subfields with each subfield observed on average 12 times. 

\begin{figure*}
    \centering
    \includegraphics[width=18cm]{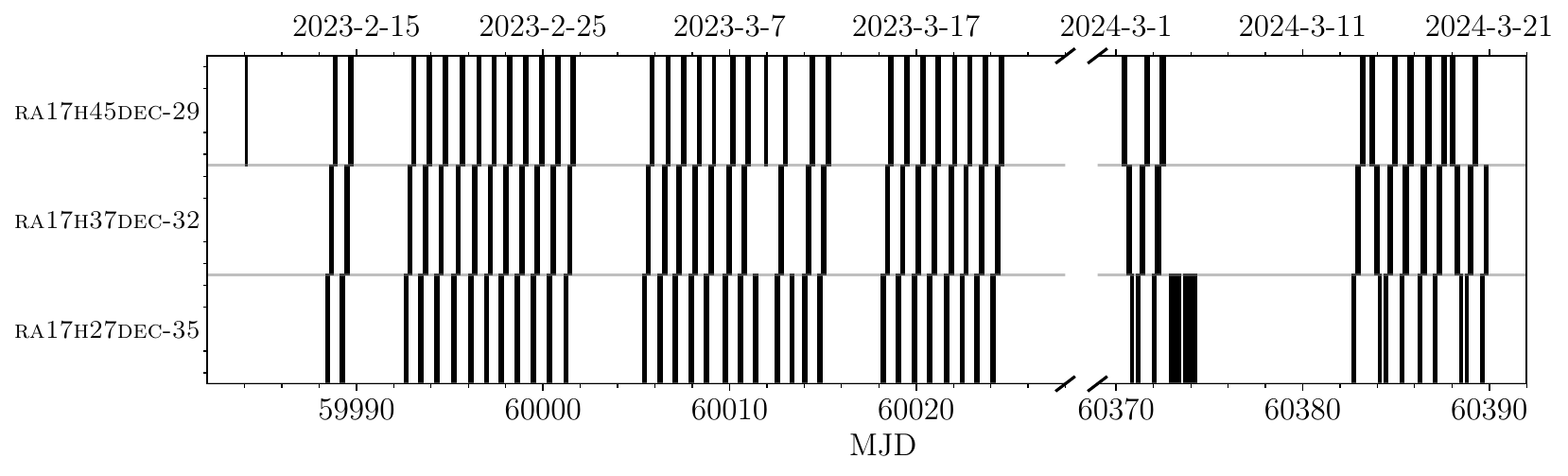}
  \caption{The observing schedule of the SPT-3G Galactic Plane Survey in 2023 and 2024. The black rectangles represent periods during which the SPT observed the given subfields, with the widths of the rectangles indicating the lengths of time spent on each. The few-day gaps between observations were dedicated to focused Sgr A\textsuperscript{*} observations. The extended coverage of the subfield \textsc{ra17h27dec-35} and the gap around MJD 60380 were due to Moon avoidance, as the Moon was close to the Galactic Plane field around that time and would have contaminated the \textsc{ra17h45dec-29} and \textsc{ra17h37dec-32} subfields.}
  \label{Figure:obs_schedule}
\end{figure*}

In 2023, we observed the Galactic Plane for 32 observing days between 2023 February 13 and 2023 March 21, and in 2024 for 13 observing days between 2024 March 1 and 2024 March 20. Some days within these periods were dedicated to focused Sgr A\textsuperscript{*} observations and not to the full Galactic Plane observations. The detailed observing schedule is shown in \autoref{Figure:obs_schedule}. The number of Galactic Plane observations conducted in each year for each subfield is reported in \autoref{Table:nobs}. Focused Sgr A\textsuperscript{*} observations and the corresponding analysis are not included in this work and will be presented in a forthcoming dedicated paper. We plan to observe the Galactic Plane field for approximately one month each year during the austral summer through the end of the SPT-3G observations, which are currently expected to continue through 2028.

\begin{table*}[ht]
\centering
\begin{tabular}{cccccccc}
\toprule
\multirow{2}{*}{Subfield} & \multirow{2}{*}{R.A. Center} & \multirow{2}{*}{R.A. Range} & \multirow{2}{*}{Decl. Center} & \multirow{2}{*}{Decl. Range} & \multicolumn{2}{c}{$N_{\text{obs,total}}/N_{\text{obs,used}}$} & \multirow{2}{*}{Total Time (h)}\\
\cmidrule(lr){6-7}
 & & & & & 2023 & 2024 & \\
\midrule
\textsc{ra17h45dec-29} & 266.25$^\circ$ & 16.40$^\circ$ & $-29^\circ$ & 3$^\circ$ & 382/321 & 162/129 & 181\\ 

\textsc{ra17h37dec-32} & 264.25$^\circ$ & 16.48$^\circ$ & $-32^\circ$ & 3$^\circ$ & 385/345 & 173/153 & 186\\ 

\textsc{ra17h27dec-35} & 261.75$^\circ$ & 16.56$^\circ$ & $-35^\circ$ & 3$^\circ$ & 411/386 & 179/167 & 196\\ 
\bottomrule
\end{tabular}
\caption{Subfield definitions with numbers of observations and cumulative time per subfield. The R.A. range and decl. range indicate boresight coverages. $N_{\text{obs}}$ refers to the number of 20-minute observations in each year. $N_{\text{obs,total}}$ represents the total number of observations in the survey before excluding noisy observations, as detailed in \autoref{Sec:observations}. $N_{\text{obs,used}}$ reflects the number of observations remaining after quality cuts. Total time refers to the number of hours the SPT observed each subfield across the two years combined. }
\label{Table:nobs}
\end{table*}

\section{Methods}
\label{Sec:observations}

\subsection{Mapmaking pipeline}

\begin{figure*}
    \centering
    \includegraphics[width=20cm]{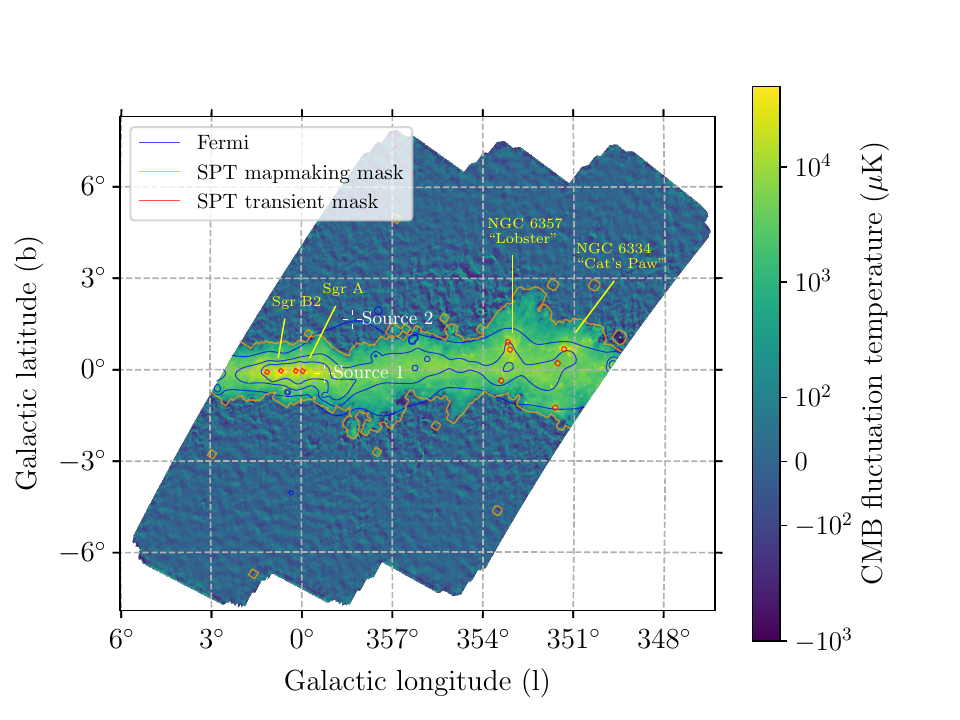}
  \caption{The SPT-3G two-year average map at 150~GHz in Galactic coordinates constructed from combining observations in 2023 and 2024. The color map employs a symmetrical logarithmic scale with a linear scale applied in the range of $\pm 100\,\mu\mathrm{K}$. This scale allows the visualization of both the bright Galactic Plane and CMB temperature fluctuations. The locations of some bright objects, including Sgr B2, Sgr A, NGC 6357 (the Lobster Nebula), and NGC 6334 (the Cat's Paw Nebula), are marked on the map. Blue lines show the contours from the Fermi Gamma-Ray Space Telescope - Large Area Telescope Galaxy map with energy range 10~GeV--2~TeV, highlighting regions of high gamma-ray emission \citep{Atwood_2009}. Orange lines indicate the SPT mapmaking mask applied during the filtering process of the mapmaking pipeline. Red lines show the SPT transient mask used in the transient detection pipeline (see \autoref{sec:transientpip}). The two white crosses mark the positions of the two transient sources described in this work. This figure is specially produced with the mapmaking mask also applied during high-pass filtering, in order to reduce ringing effects and better reveal structures in the Galactic Plane, whereas all maps used for analysis have been high-pass filtered uniformly to suppress atmospheric noise. }
\label{Figure:coadd}
\end{figure*}

To detect transients in the Galactic Plane observations, we follow a procedure similar to the one used in \cite{Guns_2021} in the SPT-3G main survey with some modifications. We construct maps for each observation using the mapmaking pipeline detailed in \cite{Dutcher_2021}. The pipeline converts the time-ordered data (TOD) to individual maps by filtering, weighting, and binning the TOD. Several filters (e.g. low- and high-pass filter, common-mode filter, and polynomial filter) are applied to the TOD to reduce noise, which mainly comes from the atmosphere. The SPT mapmaking mask shown in \autoref{Figure:coadd} is applied during the common mode filtering and polynomial filtering processes. The mask is designed to cover the full Galactic Plane, as well as bright sources outside of the plane, in order to suppress ringing from the filters and mitigate bias in the maps. After filtering, the TOD samples are binned into map pixels and averaged using inverse-variance weighting. The pixelization scheme uses the Lambert zenithal equal-area projection \citep{Calabretta_2002} with 0.25$^{\prime\prime}$ pixels.

Bad observations, including those taken during poor weather or when the telescope was not operating properly, are excluded from the analysis. In 2023, observations contaminated by the Moon were also removed. In 2024, we employed an active Moon-avoidance strategy and no data cuts for Moon contamination were required. Quality cuts are applied based on the number of functioning detectors during each observation and the variance of the resulting map. To keep as many usable observations as possible for the analysis, we apply conservative cuts, removing only clear outliers. In total, 186 observations ($\sim$11\%) are excluded from the initial set of 1,692 observations, leaving 1,506 observations for the transient analysis. 

After constructing individual maps for each valid observation, we apply pointing corrections in two steps: relative correction and absolute correction. For the relative correction, we select approximately ten bright, point-source-like sources in each subfield as reference sources. We then correct for the mean positional offset of the centroid positions of the reference sources in each observation relative to those in the first observation of the corresponding subfield. The first observation is chosen arbitrarily, as the subsequent absolute correction removes any dependence on the choice of relative pointing frame. We then generate an average map from all observations after applying the relative pointing correction. 

For the absolute correction, we use point sources from the Australia Telescope 20~GHz Survey (AT20G) catalog as the references, which have small positional uncertainties of about 0.9$^{\prime\prime}$ \citep{Murphy_2010}. Dozens of AT20G sources lie within the Galactic Plane field. 
We selected 17 bright reference sources with flux densities $>$ 50~mJy in the AT20G catalog, not located near the edges of the field, and clearly associated with a single SPT-3G counterpart in the average map. We calculate a single average positional offset by comparing the catalog positions of the reference sources to the centroid positions of their SPT-3G counterparts in the average map. This absolute offset is then corrected for in all observations uniformly. We adopt this two-step approach instead of aligning each individual map directly to the absolute AT20G frame, because there are not enough AT20G counterparts with sufficiently high signal-to-noise (S/N) in each subfield at the individual observation level. The final pointing-corrected SPT-3G maps have pointing uncertainties of a few arcseconds.


We apply a relative gain calibration by forcing the flux densities of a few static bright sources in each subfield to be the same across observations. The flux density of each source is calculated by fitting a 2D Gaussian model to the source in the map and taking the best-fit amplitude. The selection criteria for reference sources are that they should not be diffuse and their flux densities should be stable over a period of time that surpasses that of the SPT-3G Galactic Plane Survey duration. We calculate the average observed flux densities for the reference sources then determine and apply a gain conversion factor for each observation that normalizes the flux densities of reference sources in that observation to the average values. The standard deviations of the gain conversion factors are 1.4\%, 3.8\%, and 3.2\% at 95, 150, and 220~GHz, respectively. After calibration, the remaining uncertainties are 0.5\%, 1\%, and 1\% at the 95, 150, and 220~GHz bands, respectively. This relative gain calibration step helps reduce artifacts in the following differencing step. 


We make yearly average maps by computing the weighted average of all the gain-calibrated maps for each year. The weights are calculated based on the inverse variance, with pixels inside the SPT mapmaking mask excluded from the variance estimation. \autoref{Figure:coadd} shows the two-year average map for the 150 GHz~band in Galactic coordinates. The two-year map depths away from the Galactic Plane are 6.8, 8.3, and 20~$\mu\mathrm{K}$-arcmin (corresponding to 1.0, 1.6, and 4.0~mJy) in the 95, 150, and 220~GHz bands, respectively.

\begin{figure}
    \centering
    \includegraphics[width=8.5cm]{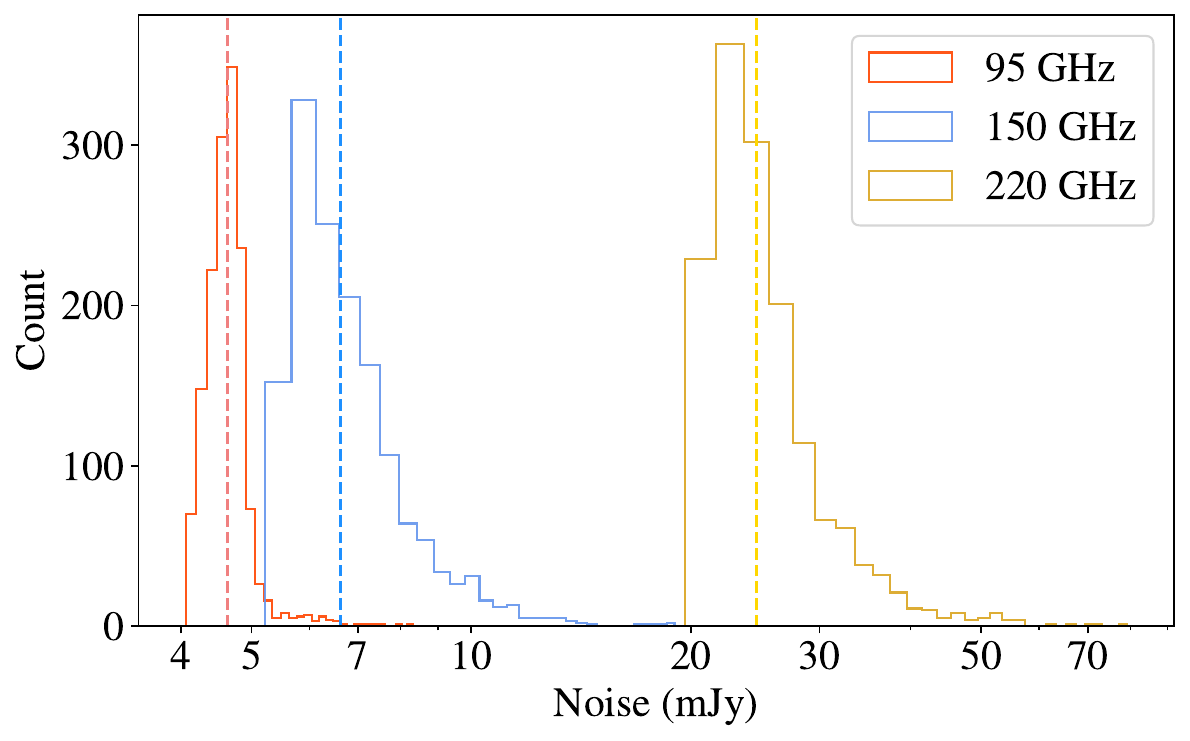}
  \caption{Histogram of the noise values for all filtered difference maps in 2023 and 2024 at 95, 150, and 220 GHz. The noise level of each individual map is estimated as the rms of a noise-dominated region located away from the Galactic Plane (signal leakage is present in the Galactic Plane). The $x$-axis is shown on a logarithmic scale. Dashed lines indicate the median noise levels of 4.6, 6.6, and 24.6~mJy at 95, 150, and 220~GHz, respectively. }
  \label{fig:noisehist}
\end{figure}

We generate a difference map from each individual observation map by subtracting the yearly average map corresponding to the year in which the observation was taken. This is to remove static backgrounds from the maps, including the CMB, emission from static sources, and dust emission from the Galaxy. While we could have constructed a single average map using all the observations from both 2023 and 2024, we choose instead to create yearly average maps. This approach better reflects the static background for each year, given the long gap of roughly 11 months between the 2023 and 2024 observations and the presence of variable sources that change on yearly timescales. Since the noise in individual maps is much higher than in the average map, we expect this choice to have a negligible impact on the transients we are able to detect. 
Ideally, this differencing step would leave only transient and variable sources in the resulting map, although there are still residuals around bright sources (see \autoref{sec:transientpip}). 
We then filter the difference maps to enhance the sensitivity to point sources as detailed in \cite{Guns_2021}. The median noise level in an individual filtered difference map is approximately 4.6, 6.6, and 24.6~mJy at 95, 150, and 220~GHz, respectively (see \autoref{fig:noisehist}). These filtered difference maps are then used as inputs to the transient detection pipeline described in \autoref{sec:transientpip}. 

\subsection{Transient detection pipeline}
\label{sec:transientpip}

Details of the transient detection pipeline can be found in \cite{Guns_2021}. The pipeline fits Gaussian flare models to lightcurves of individual pixels in the filtered difference maps, then calculates the significance of the best-fit flare model for each pixel which is later used to flag detections for further analysis. The significance is determined by the test statistic (TS) value, which is the ratio of the likelihood of the best-fit flare model to that of the null hypothesis. The TS value is roughly the square of the S/N of the event.

Some sources in the Galactic Plane are extremely bright, and even with relative gain calibration applied, bright sources can leak a significant amount of signal into the difference maps after the average map is subtracted. For instance, Sgr A has a flux density of approximately 20~Jy at 95~GHz. A 1\% leakage corresponds to 200 mJy, which, given the noise level, would appear as a 40$\sigma$ fluctuation. Regions with significant signal leakage are masked with a radius of 5$^{\prime}$ centered on the bright sources. 
To minimize the masked area, we only exclude ten sources where the leakage is bright enough to make it impossible to disentangle potential transient signals.
The transient mask is shown in \autoref{Figure:coadd}. 

To save computational time, we apply a S/N cut before performing Gaussian fits, restricting our analysis to pixels that exceed a $5\sigma$ threshold relative to local noise in both 95 and 150~GHz bands simultaneously, which we refer to as outlier pixels. Local noise is defined as the rms of the pixel values within an annulus with an inner radius of 2$^{\prime}$ and outer radius of 3$^{\prime}$ centered around each pixel. We use local noise per pixel because the noise level is not uniform across the map due to residuals from bright sources. This approach specifically addresses locally enhanced noise in the Galactic Plane caused by leaked signals that are not masked in the previous step. The size of the annulus is chosen to exclude potential signals from the center source, if present, while ensuring that the noise estimate remains relevant to the localized region.

Our implementation of the S/N cut differs from previous SPT-3G transient studies, which adopt a 3$\sigma$ threshold with noise estimated from the weight map \citep{Guns_2021,Tandoi_2024}. The more conservative threshold is due to the nature of the Galactic Plane field: unlike the extragalactic field, in which Galactic emission is low and the astrophysical noise is mostly Gaussian, the Galactic Plane field contains strong and statistically anisotropic Galactic emission, making the noise more difficult to define and characterize. 
Applying a 3$\sigma$ detection threshold in the Galactic field could introduce a large number of false detections. Therefore, we adopt a stricter 5$\sigma$ threshold for this initial study. The implementation of the S/N threshold in two bands is designed to exclude contaminants, particularly instrumental glitches, which typically appear in only one band. 
The choice to use the 95 and 150~GHz bands only is due to the fact that the noise level in the 220~GHz band is roughly 5 times higher (see \autoref{fig:noisehist}). 

We recognize that this stricter cut increases the possibility of missing real transients that fall below the threshold. The S/N cut applied to individual maps ($\sim$ 20~mins) also biases the search toward transients of certain durations, reducing sensitivity to both shorter and longer events. Should better noise characterization and false-positive identification methods be developed, future work could involve a deeper search using a $3\sigma$ threshold, single-band detections, and searches on maps with various timescales. However, these avenues are beyond the scope of the present study.


When an outlier pixel (defined here as greater than 5$\sigma$ in both 95 and 150 GHz) is detected in an observation, we construct the lightcurve for that pixel using maps from the same year and the same subfield as the observation. We then run the Gaussian flare-fitting algorithm on all outlier pixels.
We set a TS threshold of 36 in this work, which has no effect on our detections (i.e. we are effectively turning this additional cut off for this work). 
We then perform a visual inspection for all detections reported by the pipeline, excluding events such as weather balloons and leaked signals from the Galaxy. Thermal emission from weather balloons, launched as part of other research efforts at the South Pole, is a regular source of contamination. They can be easily identified by their extended shape in a single observation map, a distinct spectral index, and noticeable movement between single scan maps. Leaked signals can be identified by their proximity to bright sources and morphologies (which do not look like point sources). Asteroids may also appear in our maps, as the Galactic Plane field is close to the ecliptic plane. While no asteroids have been detected in our current search, they can be identified by their movement in the maps and have been previously studied in other SPT survey fields \citep{Chichura_2022}. 

\setlength{\tabcolsep}{3pt}
\begin{table*}[!t]
\centering
\begin{tabular}{cccccccccccc}
\toprule
\multirow{2}{*}{Source Name} & \multirow{2}{*}{R.A.}  & \multirow{2}{*}{Decl.}   & \multirow{2}{*}{$\sigma_{\mathrm{pos}}$} & \multirow{2}{*}{$l$} & \multirow{2}{*}{$b$} & Peak Time & \multicolumn{3}{c}{Peak Flux Density (mJy)} &\multirow{2}{*}{TS}\\
\cmidrule(lr){8-10}
 & & & & & & MJD & 95 GHz & 150 GHz & 220 GHz \\
\midrule
SPT-SV J174417.2-293942 & 266.072$^{\circ}$ & $-29.662^{\circ}$ & $6.1''$ & 359.229$^{\circ}$ & $-0.134^{\circ}$ & 60023.42 & $80.3 \pm 4.7$ & $59.4 \pm 6.6$ & $49.3 \pm 25.4$ & 2227\\
SPT-SV J173508.3-292956 & 263.785$^{\circ}$ & $-29.499^{\circ}$ & $6.4''$ & 358.307$^{\circ}$ & 1.640$^{\circ}$ & 60384.71 & $58.7 \pm 4.5$  & $86.6 \pm 6.2$ & $104.4 \pm 23.1$ & 1887\\
\bottomrule
\end{tabular}
\caption{Transient events detected in the SPT-3G Galactic Plane Survey in 2023 and 2024. The source locations in equatorial coordinates (R.A. and decl.) refer to the best-fit positions measured by the SPT. The pointing uncertainties, $\sigma_{\text{pos}}$, are calculated as described in \autoref{sec:sourceassociation}. The corresponding locations in Galactic coordinates ($l$ and $b$) are also reported. The peak time corresponds to the start time of the observation when the single-observation 95~GHz flux density reaches its maximum during the flare. The peak times for 95 and 150~GHz are not necessarily the same, but they are close within a few hours. The peak flux densities for 95, 150, and 220~GHz represent the single-observation flux densities for these three bands at the peak time. The TS value, as detailed in \cite{Guns_2021}, is computed from the lightcurve during the year in which the source shows flaring activity. }
\label{table:sourcelist}
\end{table*}

\section{Results}
\label{Sec:results}

During approximately 500 hours of observations conducted in 2023 and 2024, we identified two transient sources---SPT-SV J174417.2-293942 and SPT-SV J173508.3-292956---with at least one S/N $> 5 $ detection in both the 95 and 150~GHz maps. We refer to them as Source 1 and Source 2, respectively. The locations, positional uncertainties, peak time, peak flux densities, and TS values of the two detected sources are listed in \autoref{table:sourcelist}. 

\begin{figure*}
    \centering
    \includegraphics[width=18cm]{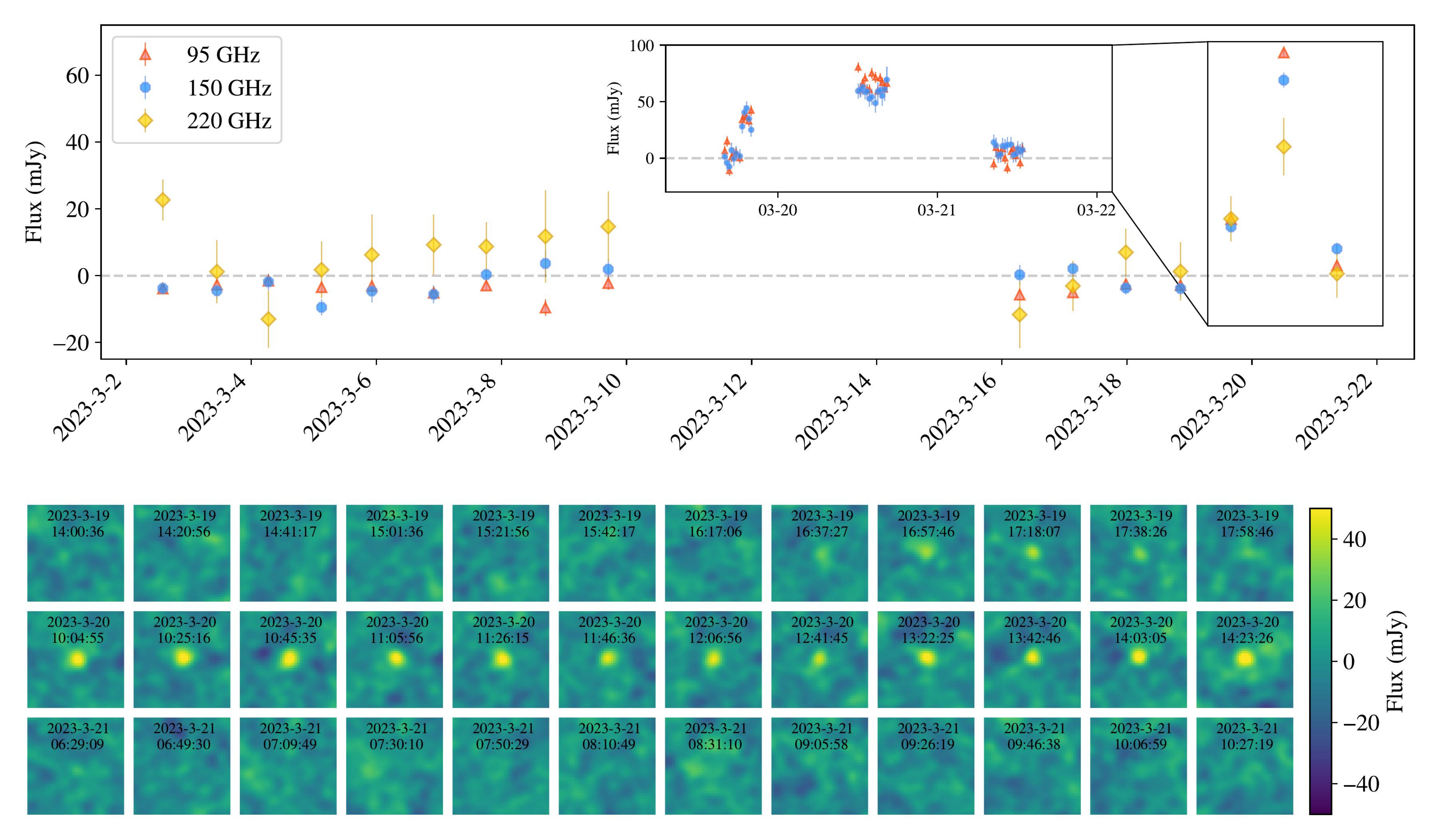}
  \caption{Lightcurve and thumbnails for SPT-SV J174417.2-293942 (Source 1). The top main panel shows the daily-averaged lightcurve for all three bands, with red triangles representing the 95~GHz band, blue circles representing the 150~GHz band, and yellow diamonds representing the 220~GHz band. The dotted gray line indicates the 0~mJy level for clarity. The inset panel zooms in on the period ranging from one day before to one day after the flares, showing the single-observation lightcurves at the 95 and 150~GHz bands. The flux densities shown here represent deviations from the observed quiescent values, as the lightcurves are extracted from difference maps. The bottom panel displays single-observation thumbnails centered on the location of Source 1 at 150~GHz for observations within the three-day window around the flare. The time labeled on each thumbnail corresponds to the start time of that observation in Coordinated Universal Time (UTC). Each thumbnail box is 10.25$^{\prime}$ $\times$ 10.25$^{\prime}$ in size.  }
  \label{fig:lc1}
\end{figure*}

\begin{figure*}
    \centering
    \includegraphics[width=18cm]{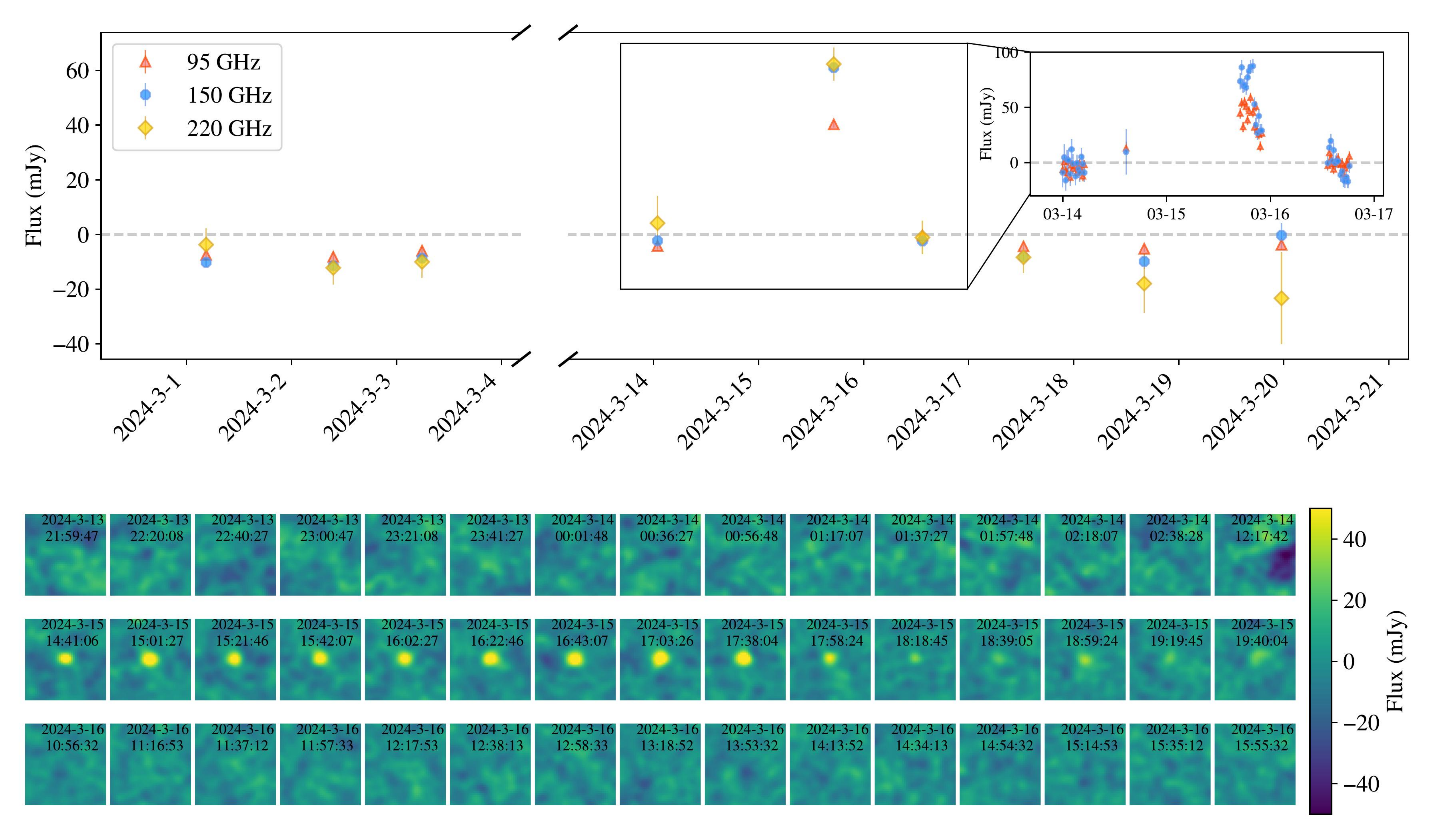}
  \caption{Lightcurve and thumbnails for SPT-SV J173508.3-292956 (Source 2). The plotting scheme is the same as that used in \autoref{fig:lc1}.}
  \label{fig:lc2}
\end{figure*}

\subsection{The SPT-3G lightcurves and thumbnails}

After detecting the two transient sources, we extracted the lightcurves for both sources around the time of their flares, presented in \autoref{fig:lc1} and \autoref{fig:lc2}. For both plots, the top main panels show daily-averaged lightcurves for all three bands, with each point representing the average flux density of the source for all observations conducted on the same observing day. For both sources, there are gaps of several days in the lightcurves; for example, between 2023 March 10 and 2023 March 16 for Source 1 and between 2024 March 4 and 2024 March 2024 for Source 2. The gaps correspond to periods when focused Sgr A\textsuperscript{*} observations were conducted. For Source 1, there is only one group of points after the flare, since the SPT switched back to main field observations shortly after the detection of the flare. For both sources, we zoom in on the periods one day before and one day after the flares and display the single-observation lightcurves for those periods in the inset plots. The single-observation lightcurves show only the 95 and 150~GHz bands, as the 220~GHz band is particularly noisy for single observations. 

Both sources are undetected in the yearly average maps. Because the noise levels at the positions of the two sources differ significantly, we compare their quiescent flux densities with the local noise levels (see \autoref{sec:transientpip}). For both sources, the measured quiescent flux densities are consistent with the local noise. We report corresponding 3$\sigma$ upper limits in \autoref{tab:catalog_sources}.



The bottom panels show single-observation thumbnails of the sources during the period ranging from one day before to one day after the flares. Each row corresponds to observations from different days: the top row corresponds to observations from the day before the flares, the middle row corresponds to observations from the day of the flares, and the bottom row corresponds to observations from the day after the flares. For Source 1, our observations capture part of the rise of the flare the day before the peak. For Source 2, observations on the day of the peak emission capture part of the fall of that flare.

We estimate the timescale of each flare by fitting a multi-band Gaussian model to the 95 and 150~GHz single-observation lightcurves simultaneously. This model assumes a common peak time and flare width across both bands, while allowing the amplitude to vary independently. Source 1 has a timescale of 1.00~$\pm$~0.05~days and Source 2 has a timescale of 0.80~$\pm$~0.08~days. These timescales are longer than those typical of mm-wave stellar flares, which usually range from minutes to hours \citep{Tandoi_2024}.

\subsection{Polarization fraction}

In addition to analyzing the temperature maps (Stokes I maps), we also measure the linear polarization fraction for the two transient events. All Stokes I/Q/U maps are constructed and filtered in the same way as described in \autoref{Sec:observations}. We can estimate the polarization fraction using the equation
\begin{equation}
\sqrt{\langle p^2 \rangle} = \sqrt{\frac{\langle Q^2+U^2\rangle}{\langle I^2\rangle}}.
\end{equation}
The I/Q/U values are taken to be the pixel values of the filtered I/Q/U maps at the source location. We independently calculate the polarization fractions for the 95, 150, and 220~GHz bands. Polarization fractions for the two transient events are calculated only during their flares, as their I/Q/U values are all close to 0 when the sources are in quiescent periods, which results in polarization fraction values that are dominated by noise and statistically unreliable. We stack all I/Q/U maps from observations conducted on the day of the flares to reduce the noise in the polarization maps. We obtain the I/Q/U values from the stacked maps and calculate the polarization fraction values. Results for both sources are statistically consistent with non-detection in polarization; therefore, we report here only the 95\% upper limit polarization fractions, as shown in \autoref{table: pol}.

\setlength{\tabcolsep}{3pt}
\begin{table}[H]
\centering
\begin{tabular}{cccccc}
\hline
\multirow{2}{*}{ID} & \multirow{2}{*}{$\alpha^{95}_{150}$} & \multirow{2}{*}{$\alpha^{150}_{220}$} & \multicolumn{3}{c}{95\% upper limit $\sqrt{\langle p^2 \rangle}$} \\
\cline{4-6}
 & & & 95~GHz & 150~GHz & 220~GHz \\
\hline
1 & $0.31\pm0.08$ & $-0.72\pm0.45$ & $<$0.10 & $<$0.10 & $<$0.41 \\
2 & $0.96\pm0.08$ & $0.06\pm0.27$ & $<$0.09 & $<$0.08 & $<$0.33 \\
\hline
\end{tabular}
\caption{Average spectral indices and 95\% upper-limit polarization fractions for the two transient events detected in the SPT-3G Galactic Plane Survey. Spectral indices and polarization fractions are calculated during the flares by averaging over observations conducted on the same day as each flare. Both events are statistically consistent with no polarization detection; thus, only the 95\% upper limits are reported here.}
\label{table: pol}
\end{table}

\begin{figure*}
    \centering
    \includegraphics[width=18cm]{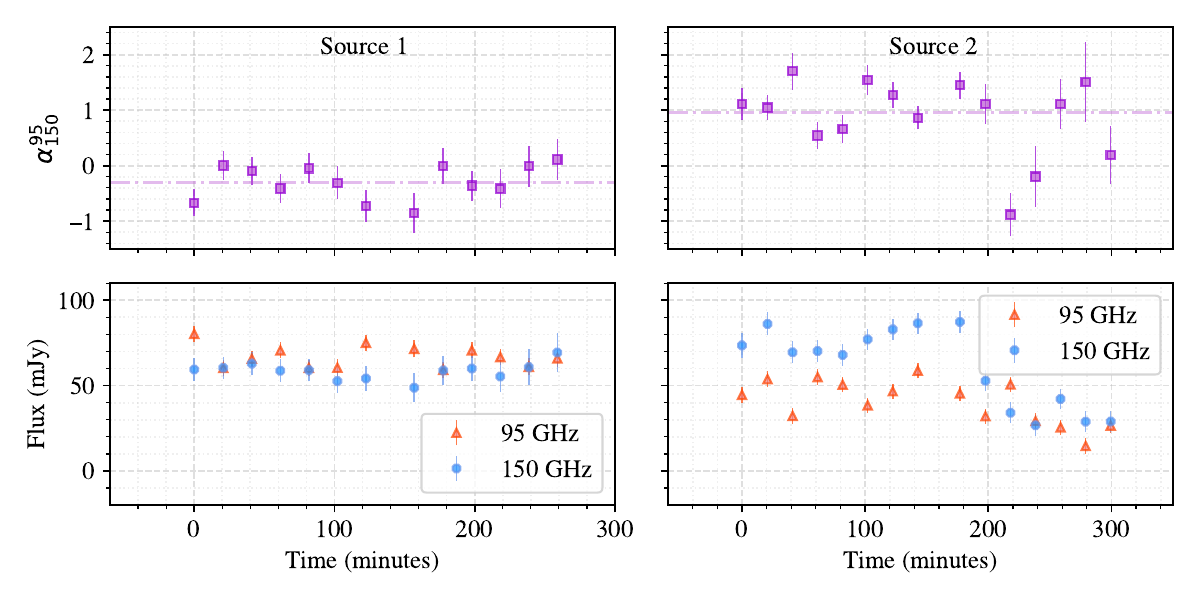}
  \caption{Spectral index evolution with corresponding lightcurves for Source 1 (left panels) and Source 2 (right panels). The top panels display the spectral index $\alpha^{95}_{150}$ for each observation during the flare period. The purple dashed line marks the average spectral index during the flare. The bottom panels show the corresponding lightcurves at 95 GHz and 150 GHz. The time indicated here is relative to the first observation on the day of the flare in minutes.}
  \label{fig:specind}
\end{figure*}

\subsection{Spectral index}

Since the SPT observes at three bands simultaneously, we calculate two separate spectral indices, $\alpha^{95}_{150}$ and $\alpha^{150}_{220}$, using the following equations:


\begin{align}
\alpha^{95}_{150} &= \frac{\text{log}(S_{95}/S_{150})}{\text{log}(\nu_{95}/\nu_{150})}  \\
\alpha^{150}_{220} &= \frac{\text{log}(S_{150}/S_{220})}{\text{log}(\nu_{150}/\nu_{220})} ,
\end{align}

\noindent where $S_{95}$, $S_{150}$, and $S_{220}$ represent the flux densities at 95, 150, and 220~GHz, respectively; $\nu_{95}$, $\nu_{150}$, and $\nu_{220}$ represent the central frequency for each band, where $\nu_{95} = 94.2~\text{GHz}$, $\nu_{150} = 147.8~\text{GHz}$, and $\nu_{220} = 220.7~\text{GHz}$; $\alpha^{95}_{150}$ represents the spectral index between 95 and 150~GHz; and $\alpha^{150}_{220}$ represents the spectral index between 150 and 220~GHz. We calculate the average spectral indices by stacking all observations conducted on the day of the flare and deriving the spectral indices using the average flux densities in the stacked map. For Source 1, the average spectral indices are $\alpha^{95}_{150} = -0.31~\pm~0.08$ and $\alpha^{150}_{220} = -0.72~\pm~0.45$. For Source 2, the average spectral indices are $\alpha^{95}_{150} = 0.96~\pm~0.08$ and $\alpha^{150}_{220} = 0.06~\pm~0.27$ (see \autoref{table: pol}). In general, Source 1 exhibits a falling spectrum while Source 2 exhibits a rising to flat spectrum. We also calculate the spectral indices from each observation during the flare time, which are shown in \autoref{fig:specind}. Only $\alpha^{95}_{150}$ is included in \autoref{fig:specind}, since $S_{220}$ has large error bars at the single-observation level, resulting in error bars significantly larger than the measured $\alpha^{150}_{220}$ values.

\setlength{\tabcolsep}{6pt}
\begin{table*}[!t]
\hspace*{-1cm}
\centering
\begin{tabular}{cccc}
\hline
\textbf{Catalog} & \textbf{Catalog identifier} & \textbf{Band} & \textbf{Photometry}  \\
\hline
\multicolumn{4}{c}{\textbf{Source 1}} \\
\hline
3XMM-DR8 & J174417.2-293944 & Band 8 (0.2--12.0~keV) & $3.74~\times~10^{-13}~\text{erg}~\text{cm}^{-2}~\text{s}^{-1}$ \\
CSC 2.1 & 2CXO J174417.2-293944 & HRC wide (0.1--10.0~keV) & $8.69~\times~10^{-14}~\text{erg}~\text{cm}^{-2}~\text{s}^{-1}$\\
&  & ACIS broad (0.5--7.0~keV) & $2.77~\times~10^{-13}~\text{erg}~\text{cm}^{-2}~\text{s}^{-1}$ \\
&  & ACIS hard (2.0--7.0~keV) & $2.49~\times~10^{-13}~\text{erg}~\text{cm}^{-2}~\text{s}^{-1}$ \\
&  & ACIS medium (1.2--2.0~keV) & $1.18~\times~10^{-13}~\text{erg}~\text{cm}^{-2}~\text{s}^{-1}$ \\
&  & ACIS soft (0.5--1.2~keV) & $6.84~\times~10^{-14}~\text{erg}~\text{cm}^{-2}~\text{s}^{-1}$ \\
&  & ACIS ultrasoft (0.2--0.5~keV) & $1.22~\times~10^{-14}~\text{erg}~\text{cm}^{-2}~\text{s}^{-1}$ \\
\textit{Gaia} DR3 & 4057051396569058432 & $G$ (350--1000~nm) & 12.73 mag \\
& & $G_{\text{BP}}$ (330--680~nm) & 13.82 mag \\
& & $G_{\text{RP}}$ (640-1000~nm) & 11.70 mag \\
& & $G_{\text{RVS}}$ (847-874~nm) & 11.26 mag \\
2MASS & 17441724-2939444 & J (1.24~$\mu$m) & 10.21 mag \\
& & H (1.66~$\mu$m) & 9.30 mag \\
& & Ks (2.16~$\mu$m) & 9.01 mag \\
All\textit{WISE} & J174417.30-293946.4 & W1 (3.4~$\mu$m) & 7.68 mag \\
&  & W2 (4.6~$\mu$m) & 7.38 mag \\
&  & W3 (12~$\mu$m) & 7.92 mag \\
&  & W4 (22~$\mu$m) & 5.52 mag \\
SPT-3G & SPT-SV J174417.2-293942 & 220~GHz (1.4~mm) & $<$2,760~mJy \\
(this work)&  & 150~GHz (2.0~mm) & $<$581~mJy \\
&  & 95~GHz (3.2~mm) & $<$408~mJy \\
VAST & N/A & 888~MHz (0.3~m) & $<$2.82~mJy \\
\hline
\multicolumn{4}{c}{\textbf{Source 2}} \\
\hline
CSC 2.1 & 2CXO J173508.2-292957 & ACIS broad (0.5--7.0~keV) & $6.93~\times~10^{-13}~\text{erg}~\text{cm}^{-2}~\text{s}^{-1}$\\
&  & ACIS hard (2.0--7.0~keV) & $4.44~\times~10^{-13}~\text{erg}~\text{cm}^{-2}~\text{s}^{-1}$ \\
&  & ACIS medium (1.2--2.0~keV) & $1.46~\times~10^{-13}~\text{erg}~\text{cm}^{-2}~\text{s}^{-1}$ \\
&  & ACIS soft (0.5--1.2~keV) & $1.15~\times~10^{-13}~\text{erg}~\text{cm}^{-2}~\text{s}^{-1}$ \\
\textit{Gaia} DR3 & 4058581921170339456 & $G$ (350--1000~nm) & 11.13 mag  \\
& & $G_{\text{BP}}$ (330--680~nm) & 12.14 mag \\
& & $G_{\text{RP}}$ (640-1000~nm) & 10.14 mag \\
2MASS & 	17350831-2929580 & J (1.24~$\mu$m) & 8.51 mag \\
& & H (1.66~$\mu$m) & 7.76 mag \\
& & Ks (2.16~$\mu$m) & 7.44 mag \\
All\textit{WISE} & J173508.32-292958.0 & W1 (3.4~$\mu$m) & 7.05 mag  \\
&  & W2 (4.6~$\mu$m) & 7.28 mag  \\
&  & W3 (12~$\mu$m) & 7.26 mag \\
&  & W4 (22~$\mu$m) & 7.34 mag  \\
SPT-3G & SPT-SV J173508.3-292956 & 220~GHz (1.4~mm) & $<$71~mJy \\
(this work) &  & 150~GHz (2.0~mm) & $<$18~mJy \\
&  & 95~GHz (3.2~mm) & $<$9~mJy \\
VAST & N/A & 888~MHz (0.3~m) & $<$0.23~mJy \\

\hline
\end{tabular}
\caption {Catalog associations for the two sources, along with corresponding source names, observing bands, and quiescent flux density (photometry) measurements. The flux densities from the CSC included here are from the \texttt{flux\_aper\_avg} entry in the original catalog, which represent the aperture-corrected net energy flux inferred from the source region aperture, averaged over all contributing observations, and calculated by counting X-ray events \citep{Evans_2024}. Photometry from \textit{Gaia}, 2MASS, and \textit{WISE} is expressed in the Vega magnitude system. Both sources are not detected in the SPT-3G yearly average maps and VAST mosaic map. Therefore, we report only the 3$\sigma$ upper limits. }
\label{tab:catalog_sources}
\end{table*}

\subsection{External catalog association}
\label{sec:sourceassociation}

The SPT positions are the centroids of the TS maps of each event to maximize localization precision \citep{Guns_2021, Tandoi_2024}. The SPT pointing uncertainties in each axis for transient events are calculated using the following equations:

\begin{align}
\sigma^{2}_{\mathrm{xdecl.}} &= \sigma_{\mathrm{xdecl.,sys,abs}}^{2} + \sigma_{\mathrm{xdecl.,sys,rel}}^{2}+\left(\frac{\theta_{\mathrm{beam,xdecl.}}}{\sqrt{\text{TS}}}\right)^{2} \\
\sigma^{2}_{\mathrm{decl.}} &= \sigma_{\mathrm{decl.,sys,abs}}^{2} + \sigma_{\mathrm{decl.,sys,rel}}^{2}+\left(\frac{\theta_{\mathrm{beam,decl.}}}{\sqrt{\text{TS}}}\right)^{2},
\end{align}

where xdecl. is defined as $\text{R.A.}\cdot\cos({\text{decl.}})$ to account for the fact that the distance corresponding to a given R.A. interval depends on declination. The first term represents the systematic absolute pointing uncertainty, the second term represents the systematic relative pointing uncertainty, and the third term represents the statistical pointing uncertainty. The first term is calculated as the standard deviation of the positional offsets between AT20G reference sources and their counterparts in the pointing-corrected average map, where $\sigma_{\mathrm{xdecl.,sys,abs}} = 3.05^{\prime\prime}$ and $\sigma_{\mathrm{decl.,sys,abs}} = 4.42^{\prime\prime}$. The second term is estimated separately for the two different events as the mean of the standard deviations of the positional offsets of reference sources used for relative pointing, computed between observation maps during the flare of the event and the average map. For Source 1, $\sigma_{\mathrm{xdecl.,sys,rel}} =1.67^{\prime\prime}$ and $\sigma_{\mathrm{decl.,sys,rel}} = 1.94^{\prime\prime}$. For Source 2, $\sigma_{\mathrm{xdecl.,sys,rel}} = 2.24^{\prime\prime}$ and $\sigma_{\mathrm{decl.,sys,rel}} = 2.05^{\prime\prime}$. The third term corresponds to the ratio of the beam width to S/N of the event, where $\theta_{\mathrm{beam,xdecl.}} = 50^{\prime\prime}$ and $\theta_{\mathrm{beam,decl.}} = 49^{\prime\prime}$ \citep{Tandoi_2024}. The pointing uncertainties are reported in \autoref{table:sourcelist}.

The total positional uncertainty, combining contributions from both axes, can then be calculated using the following equation:

\begin{equation}
\sigma_{\text{pos}}^2 = \sigma_{\text{xdecl.}}^{2}+\sigma_{\text{decl.}}^{2}.
\end{equation}

\noindent The positional uncertainty, $\sigma_{\text{pos}}$, is $6.1''$ for Source 1 and $6.4''$ for Source 2 (\autoref{table:sourcelist}). 

\begin{figure*}
    \centering
    \includegraphics[width=18cm]{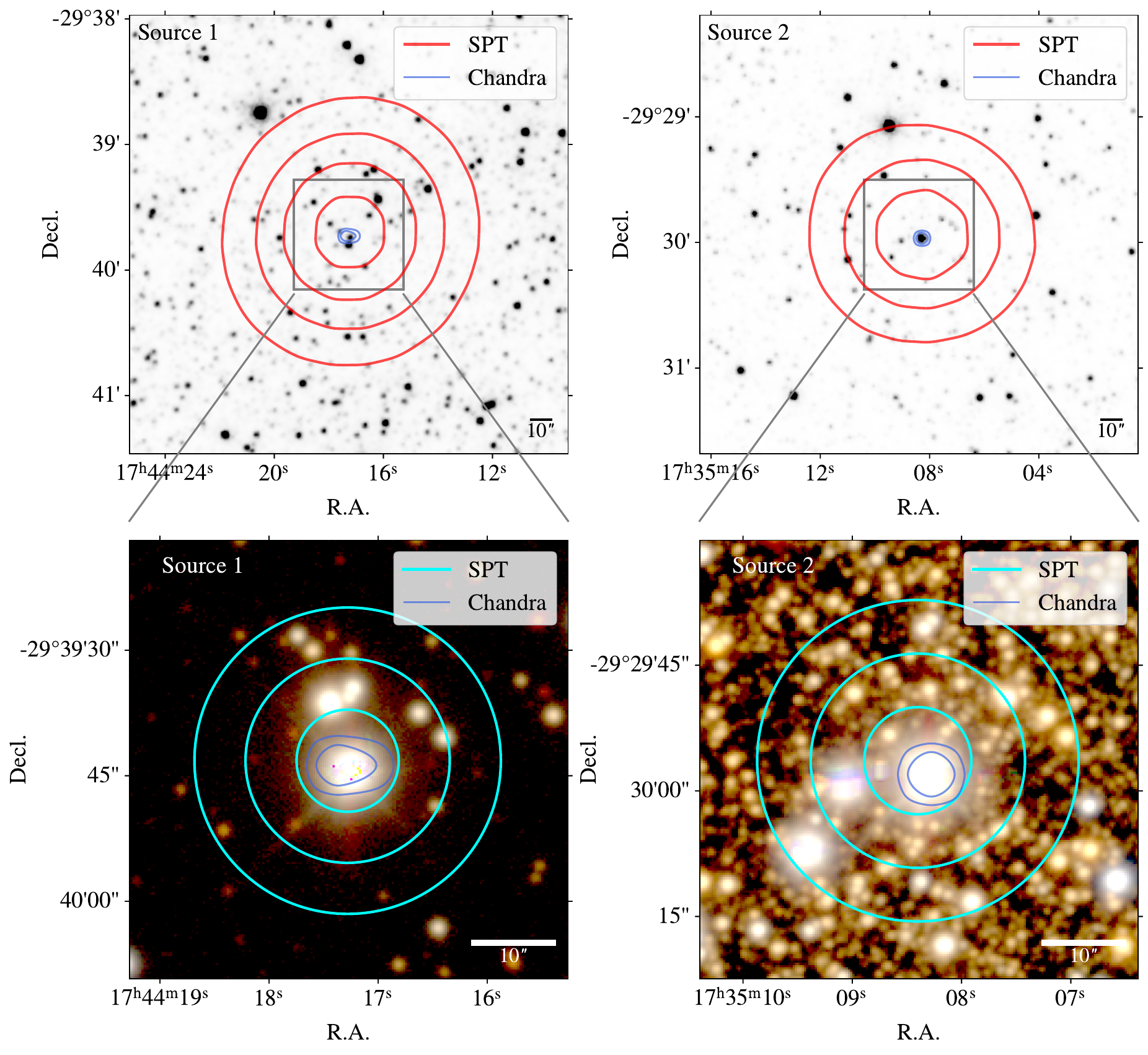}
  \caption{\textit{Spitzer}/IRAC(top) and DECaPS2(bottom) images centered on the SPT positions of the two detected transient sources. Top panels: grayscale images from IRAC channel 1 (3.6~$\mu\text{m}$) of the \textit{Spitzer}/IRAC Survey of the Galactic Center \citep{irac} display surface brightness with the color scale linearly spanning from 0 to 500 MJy/sr. Red contours show the SPT-3G TS maps at 500, 1000, 1500 and 2000 levels when available, smoothed with a 10$\sigma$ Gaussian filter.  Blue contours show the \textit{Chandra} photon count at 10 and 30$\sigma$ levels, smoothed with a 2$\sigma$ Gaussian filter. Top left panel: \textit{Chandra} contours are generated from observation\dataset [\textit{Chandra} ObsId 2278]{https://doi.org/10.25574/02278}. Top right panel: \textit{Chandra} contours are generated from observation\dataset[\textit{Chandra} ObsId 9997]{https://doi.org/10.25574/09997}. Bottom panels: Three-color DECaPS2 images are generated with the \textit{z}, \textit{i}, and \textit{r} bands using a logarithmic stretch from 0.04 to 5.00 Analog-Digital Units in all three bands. The blue \textit{Chandra} contours are the same as those in the top panels. The cyan contours are centered on the SPT source positions with radii corresponding to 1, 2, and 3$\sigma_{\text{pos}}$.}
  
  \label{fig:association}
\end{figure*}

Both transient events are located within the observing fields of the \textit{Chandra} Galactic Bulge Survey (GBS, \citealt{Jonker_2011}; \citealt{Jonker_2014}), which provides X-ray coverage for these fields. We identify clear counterparts for both sources in the \textit{Chandra} source catalog \citep{Evans_2010, Evans_2020, Evans_2024}. The angular separation between Source 1 and 2CXO~J174417.2-293944 is $1.19''$. The angular separation between Source 2 and 2CXO~J173508.2-292957 is $2.09''$. For both sources, the positional offsets lie within $\sigma_{\text{pos}}$. Source associations with \textit{Chandra} are shown in \autoref{fig:association}, where the SPT and \textit{Chandra} contours are overlaid on the images from the \textit{Spitzer}/IRAC Survey of the Galactic Center \citep{irac} and the Dark Energy Camera Plane Survey 2 (DECaPS2; \citealt{DECaPS2}). 

Using a maximum association radius of $3\sigma_{\text{pos}}$, and given that there are 1,640 X-ray sources in the \textit{Chandra} catalog from GBS covering 12 deg$^{2}$ (corresponding to a source density of 0.04 sources per square arcminute) \citep{Jonker_2011, Jonker_2014}, we estimate that there is at most a 1\% chance that a random X-ray source would be falsely associated with an SPT transient event. Since the likelihood of a false association with \textit{Chandra} is low, we conclude that our association with \textit{Chandra} is secure. 

We also perform an independent cross-match to the \textit{Gaia} DR3 source catalog \citep{gaiamission,GaiaDR3} using the p-value method described in \cite{Tandoi_2024}. For Source 1, the best match with the lowest p-value is DR3~4057051396569058432. For Source 2, the best match with the lowest p-value is DR3~4058581921170339456. Both associations have p-values less than 0.02. The \textit{Gaia} counterparts are consistent with the \textit{Chandra} associations.

We then use the \textit{Chandra} positions to search in the VizieR database \citep{Ochsenbein_2000}, since \textit{Chandra} provides smaller positional uncertainties---0.71$^{\prime\prime}$ in each axis for 95\% confidence---compared to the SPT \citep{Evans_2024}. In \autoref{tab:catalog_sources}, we list counterparts for both sources, along with quiescent flux density measurements in corresponding bands, from the 3XMM-DR8 catalog \citep{Rosen_2016}, the \textit{Chandra} Source Catalog Release 2 (CSC 2.1; \citealt{Evans_2024}), the \textit{Gaia} DR3 \citep{gaiamission,GaiaDR3}, the Two Micron All Sky Survey (2MASS) point source catalog \citep{Skrutskie_2006}, and the All\textit{WISE} source catalog \citep{Wright_2010}, where available.

We assembled observations covering the SPT Galactic region taken as part of the VAST Pilot Survey \citep{Murphy_2021} from the CSIRO ASKAP Science Data Archive (CASDA), including observations from nine epochs in 2019 and 2020.  Each observation was approximately 12\,min in duration with rms noise levels of 0.25\,mJy away from the Galactic plane increasing to 0.5–1\,mJy at $b=0\degr$, depending on the Galactic latitude.  We averaged all observations using \texttt{swarp} \citep{Bertin_2002} and inverse variance weighting. The measured quiescent flux densities at both source locations are consistent with the local noise, indicating non-detection. We report 3$\sigma$ upper limits in \autoref{tab:catalog_sources}.

Source 1 is detected and extensively studied in the Swift Bulge Survey, where it is argued to be a CV with a subgiant companion \citep{Shaw_2020}. \cite{Munari_2021} conducted photometric and spectroscopic monitoring of Source 2, concluding that it is a symbiotic star with a K4III giant companion.

\begin{figure}
    \hspace*{-1cm}
    \centering
    \includegraphics[width=9.5cm]{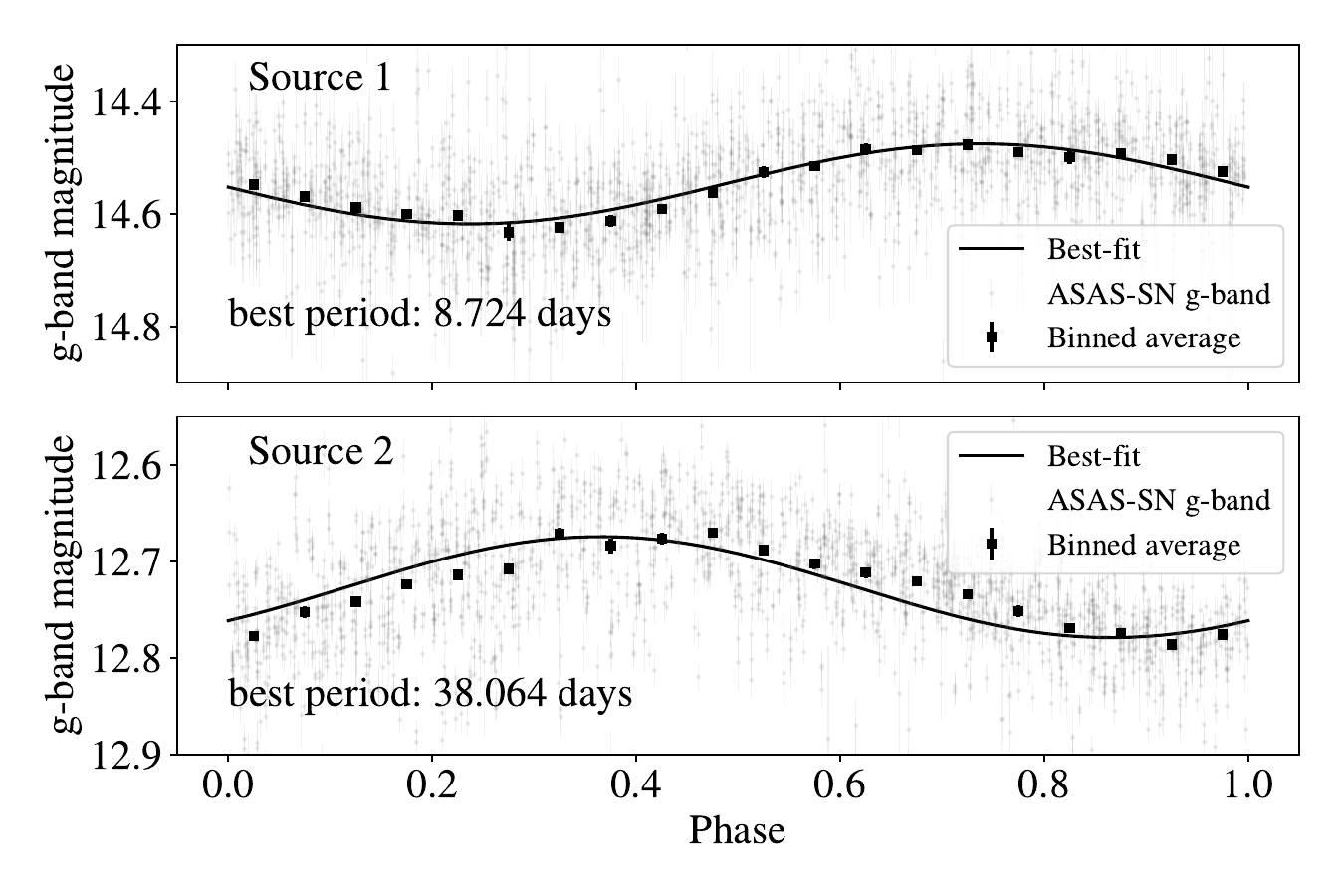}
  \caption{ASAS-SN $g$-band lightcurves folded on the optimal periods for each source as inferred from the Lomb-Scargle periodograms. Top panel: ASAS-SN lightcurve for Source 1 folded on a periodicity of 8.72~days. Bottom panel: ASAS-SN lightcurve for Source 2 folded on a periodicity of 38.04~days. Gray points show the ASAS-SN g-band magnitude measurements from MJD 59000 to the present. Black squares are phase-binned averages and the black curves are the best-fit folded lightcurves. }
  \label{fig:asassn}
\end{figure}

To investigate the long-term optical activity of the sources, we queried data from the All-Sky Automated Survey for SuperNovae (ASAS-SN -- see \citealt{Shappee_2014, Kochanek_2017}), which provides optical lightcurves for sources down to 18$^{th}$ magnitude with near-daily cadence. We perform a Lomb-Scargle periodogram analysis on the $g$-band lightcurves from MJD 59000 to the present to search for periodicities that may indicate potential binary orbital periods \citep{lomb_1976,scargle_1982}. For Source 1, the optimal period with the highest power is 8.72~days with 204 cycles included in the analysis. This result is consistent with the 8.69~day period reported by ASAS-SN, the 8.69~$\pm$~0.05 day period from a Lomb-Scargle analysis reported by \cite{Shaw_2020}, and the 8.7092~$\pm$~0.0048 day orbital period derived from radial velocity measurements in the same work \citep{Shaw_2020}. For Source 2, the optimal period is 38.04~days with $\sim$47 cycles included. \cite{Munari_2021} also report a period of 38~days from the $V$-band lightcurve of this source, although they do not draw firm conclusions about the orbital period due to a lack of radial velocity measurements.  The false alarm probabilities for both sources' periods are low (below $10^{-10}$), indicating high statistical significance. The ASAS-SN $g$-band lightcurves, folded on the optimal periods for each source, are shown in \autoref{fig:asassn}. The ASAS-SN lightcurves show no evidence of simultaneous optical flares for either event.

\subsection{Luminosity}

Using measurements of the peak flux densities of the events and distances to the sources from the \textit{Gaia} DR3 catalog \citep{GaiaDR3}, we calculate the peak isotropic mm-wave luminosities of the events using the following equation:
 
\begin{equation}
\nu L_{\nu,\text{peak}} = 4\pi d^2 \nu F_{\nu,\text{max}},
\label{equation:luminosity}
\end{equation}

\begin{table*}[!t]
\hspace*{-2cm}
\centering
\begin{tabular}{cccccccc}
\hline
ID & \textit{Gaia} DR3 Association & d (pc) & $\sigma_{\text{d}}$ (pc) & $\nu L_{\nu}^{95}$ ($\text{erg}~\text{s}^{-1}$) & $\nu L_{\nu}^{150}$ ($\text{erg}~\text{s}^{-1}$) & $\nu L_{\nu}^{220}$ ($\text{erg}~\text{s}^{-1}$) \\
\hline
Source 1 & 4057051396569058432 & 966 & 18 & $8.54 \times 10^{30}$ & $9.95 \times 10^{30}$ & $1.21 \times 10^{31}$ \\
Source 2 & 4058581921170339456 & 1497 & 44 & $1.50 \times 10^{31}$ & $3.48 \times 10^{31}$ & $6.15 \times 10^{31}$ \\
\hline
\end{tabular}
\caption{Estimated luminosities in the SPT-3G bands for the two detected transient events. Distances (d) and their uncertainties ($\sigma_{\text{d}}$) are derived from the parallaxes reported in the \textit{Gaia} DR3 catalog \citep{GaiaDR3}. The mm-wave luminosity $\nu L_{\nu}$ is calculated following \autoref{equation:luminosity}. }
\label{table:energy}
\end{table*}

\begin{figure}
    \hspace*{-1cm}
    \centering
    \includegraphics[width=10.5cm]{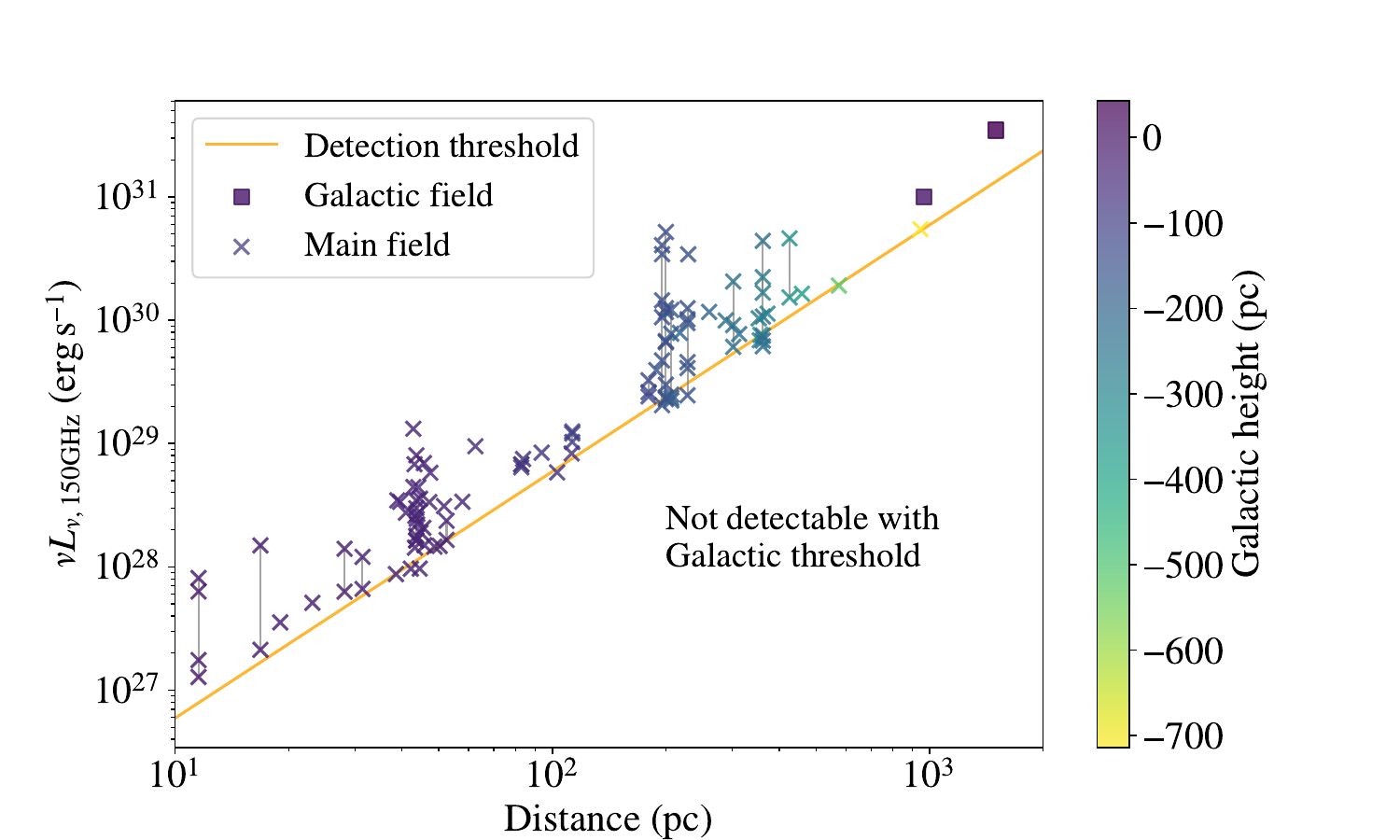}
  \caption{Peak flare luminosity versus distance at 150~GHz for the two flares reported in this paper (squares) and flares in the main SPT-3G survey (crosses). Point color indicates the Galactic height, which is the vertical distance from the Galactic Plane. Gray lines connect flares originating from the same source. The orange line represents the expected 5$\sigma$ detection threshold of the Galactic survey estimated from the average noise level at 150~GHz.   }
  \label{fig:lum_compare}
\end{figure}

\noindent where $\nu L_{\nu,\text{peak}}$ is the peak isotropic luminosity, $\nu$ is the observed frequency, $d$ is the distance to the source, and $F_{\nu,\text{max}}$ is the peak flux density at the observed frequency. The results are presented in \autoref{table:energy} and the luminosities range from $8~\times~10^{30}~\text{erg}~\text{s}^{-1}$ to $7~\times~10^{31}~\text{erg}~\text{s}^{-1}$ in the SPT-3G bands. Compared to SPT-3G transient sources reported in \cite{Guns_2021} and \cite{Tandoi_2024}, which have peak luminosities at 150~GHz ranging from $10^{26}$ to $10^{31}~\text{erg}~\text{s}^{-1}$, the two Galactic transient sources lie at the higher end of this range. \autoref{fig:lum_compare} shows the comparison to \cite{Tandoi_2024}.

\section{Discussion}
\label{Sec:discussion}

The emission mechanism for the two detected transient events remains unclear. The measured spectral indices are suggestive of a scenario in which both flares are driven by self-absorbed synchrotron emission, with their different spectral indices arising from variations in optical depth \citep{Katz_2017,Benz_2010}. The weak polarization signal with large uncertainty does not provide meaningful constraints on the emission mechanism. We consider three possible sources of emission: the white dwarf itself, the donor star, and the accretion disk, each of which is discussed in detail in the following sections.  

\subsection{Emission from the white dwarf itself?}

It is unlikely that the white dwarf itself is the source of the emission. Producing the observed flux densities within the white dwarf region would require coherent emission processes, as the brightness temperature estimated from \autoref{equation:tb} would be higher than $10^{12}$~K, which is not consistent with synchrotron emission. Moreover, the detections appear to favor long-period accreting white dwarf systems. CVs typically have orbital periods of only a few hours \citep{Schaefer2024}, while symbiotic stars, though they have longer orbital periods ranging from months to years \citep{munari_2019}, are significantly rarer than CVs (see \autoref{sec:dis-accdisk} for population size estimates). Given that short-period accreting white dwarf systems are more common, if the white dwarf were the source of the emission, we would expect more detections associated with short-period systems---which we do not observe. The long-period accreting white dwarf systems host larger donor stars and accretion disks compared to those in short-period systems. Thus, we infer that the flare likely originates from the accretion disk or the donor star, rather than from the white dwarf itself. 

\subsection{Emission from the donor star or accretion disk?}
\label{sec:dis-accdisk}

Millimeter flaring has been seen from evolved stars, like the donor stars in these two accreting white dwarf systems, in the past \citep{Kowalski_2024}.  A first question to address is whether the flares are likely to be due to the donor stars with the associated accreting white dwarfs being coincidental.  If the presence of an accreting white dwarf substantially enhances the likelihood of millimeter flares, this will give credence to models that allow production of flares from the accretion flow, rather than from the star.

We can assess this likelihood by comparing the population sizes of accreting white dwarf systems, such as those we detect, with systems containing a similar donor star but no white dwarf. The control group we use for Source 1 (a CV) is RS CVn stars, which are binary systems with the primary being a subgiant or giant star and the secondary being a dwarf or subgiant \citep{RS_CVn}. For Source 2 (a symbiotic star), we choose the control group to be coronally active red giants.

Long-period CVs are sufficiently rare that their numbers are not well-characterized.  Within 150~pc, CV populations are nearly complete.  No CVs are seen with orbital periods longer than one day and the incompleteness is expected to be almost exclusively of faint objects at shorter periods \citep{Pala}. The overall space density of CVs was found to be $4.8^{+0.6}_{-0.8}\times10^{-6}$~pc$^{-3}$ from a sample of 42 objects, so we can expect that the space density of long-period CVs is less than about $10^{-7}$~pc$^{-3}$ (other samples which are larger, albeit with less well-characterized selection effects, also show that about 1\% of CVs are at periods longer than a day; see \citealt{Ritter}). Taking a local stellar mass density of $0.04~M_\odot$~pc$^{-3}$, we estimate that there are $2.5\times10^{-6}$ long-period CVs per solar mass in the Milky Way.  Then, with the Milky Way's total stellar mass of $6\times10^{10} ~M_\odot$, we expect about $1.5\times10^5$ long-period CVs in the Galaxy \citep{LicquiaNewman}.  The population size for symbiotic stars is estimated to be about $5\times10^4$ in the Galaxy \citep{Laversveiler_2025}.

For RS CVn stars, estimates are that space densities are at least $10^{-6}$~pc$^{-3}$ \citep{Hall1976}, a factor of at least 10 higher than the space density of long-period CVs.  About 2\% of red giants are rapidly rotating \citep{Gaulme} and they have a space density of about $3.9\times10^{-4}$~pc$^{-3}$, so the space density of active red giants is about $7.8\times10^{-6}$~pc$^{-3}$ and the total population size is then about $4.7\times10^5$ in the Galaxy, which is about 10 times larger than the population of symbiotic stars.  We can thus estimate that in each case, there is only a 10\% chance that the flares would come from the objects in accreting white dwarf systems if such objects were equally likely to flare as generic rapidly rotating stars of the same spectral type. Thus, there is a combined probability of about 1\% that we would see two flares, both from accreting white dwarf systems. We can thus make a statistical argument that is strong, but not definitive, that the presence of accretion disks is likely relevant to the production mechanism for the flares.

\subsection{Connection to other flaring events in accreting white dwarfs?}
 
Recently, strong flares have been reported in the optical regime in several CVs \citep{Schaefer2022, Scaringi}.  These have been explained as magnetic reconnection events in the CV disk \citep{Schaefer2024}, following a model originally developed to understand flaring in the disks of young stellar objects \citep{Hayashi1996}. Alternatively, they may be magnetically confined thermonuclear runaways that burn only in a small region on the surface of a white dwarf, billed as ``micronovae" \citep{Scaringi}.  

It is thus worth determining whether these mechanisms might also explain millimeter flares from accreting white dwarf systems.  We note that we do not see optical flares in the ASAS-SN data for the two objects studied in this paper, but given that the donor stars are much more luminous than those of the CVs which have shown flaring, there may be optical flares buried under the donor stars.
 
We can estimate the brightness temperature $T_{b}$ of the mm-wave emission using \autoref{equation:tb}, taking into account inferences we can make about the actual size of the emitting regions in the two cases. For the magnetic reconnection scenario, the size should be of order of the orbital separation in the system, while for the thermonuclear explosion scenario, the size scale should be no more than 1 day times the ejecta speed (which for typical nova explosions are about 3000~km~s$^{-1}$ or less, close to the escape speeds of white dwarfs). The latter will be larger, and for distances of 1~kpc, will lead to angular sizes of about 1.5~$\mathrm{mas}$.
 
Then, we can write:
\begin{equation}
T_b = 2.7 \times 10^6\, \mathrm{K} \, \frac{I/50\, \mathrm{mJy}} {(\nu/100\, \mathrm{GHz})^2 (\theta/1.5\, \mathrm{mas})^2},
\label{equation:tb}
\end{equation}
in which $I$ is the flux density, $\nu$ is the frequency of observation, $\theta$ is the angular size scale of the emitting region, and $T_b$ is the brightness temperature.  Here, the values for $I$ and $\nu$ are approximately those for the SPT flares and the value of $\theta$ is the value expected for the micronova scenario. The accretion disk scenarios can be expected to have angular sizes $10-100$ times smaller and hence brightness temperatures in the $10^8-10^{10}$~K range. Classical novae, when emitting thermally, typically have brightness temperatures in the radio and millimeter bands that are less than about $10^5$~K \citep{Chomiuk}, and thus the observations require synchrotron emission in either scenario.  Some classical novae do show synchrotron emission due to shocks.  If the micronova phenomenon  does occur and explains these flares as well as the optical ones, shocking would be required. This may be challenging for the micronova scenario, because in that scenario, the outflows are presumably launched away from the magnetic poles rather than isotropically, and with the rotation of the white dwarf, successive episodes of outflow will be launched in different directions.

Still, the association of these flares with long-period accreting white dwarf systems, which are much rarer than standard CVs, favors the scenario of magnetic reconnection in the accretion disk, as this gives a reason for preferring long-period systems (which have larger disks and hence can produce stronger flares) over standard CVs. In the micronova scenario, the flare energetics will be determined by the size scale of the ignition region on the white dwarf's surface, which relates to the white dwarf's magnetic field properties and not to the binary orbital properties.

We emphasize that this discussion is speculative, and this line of reasoning depends on establishing in future data sets that these flares are generated in the disk and are not just extreme stellar flares from the donor stars. In at least one case, clear evidence for radio flares driven by magnetic reconnection in some CVs has been seen in the form of strong circular polarization \citep{Ridder}; this flaring has been attributed to the donor star, and in some of the cases, the accretion is magnetically channelled, so that no disk forms. If one or both of these two SPT transients are seen to be recurring, as might be expected if they are related to the strong optical flares seen in CVs, then future circular polarization measurements would be straightforward to make with radio follow-up.  

What we can say at the present time is that there are no obvious problems with a scenario in which the flaring occurs due to magnetic reconnections in an accretion disk.  There are some statistical problems with event rates if the flaring occurs in the donor stars, simply because the donor stars in these systems are rapidly rotating. The donor stars in binaries with mass transfer are rotating maximally for their masses and radii, and there are some indications that this might make them more likely to flare than the average stars which are rapidly rotating \citep{Medina_2020,olah_2021}. Still, it is likely that this is quite a minor effect, as CVs do not dominate the rates of flares of normal stars and the optical flaring rates of both M dwarfs and giants show only weak correlation with rotation rate for rapidly rotating stars \citep{Medina_2020, olah_2021}. The micronova scenario has a hard time explaining the millimeter flares for multiple physical reasons -- there is no reason for there to be a preference for such long-period objects and shocks would have to set in quite quickly.  With data sets from future millimeter surveys that cover a larger number of nearby CVs, including some shorter-period ones, it may become easier to determine if there are connections between strong optical flares and millimeter flares.

\subsection{Overall event rates}
We can compare results from this work to previous transient surveys with SPT-3G, where all identified Galactic transients were stellar flares. We have detected two flares in the two-month SPT-3G Galactic Plane Survey covering 100~$\text{deg}^2$. In a total of roughly 32 months of observation of the SPT-3G main field (covering 1500~$\text{deg}^2$ of extragalactic sky), \cite{Tandoi_2024} presented detections of a total of 111 stellar flare events using a 3$\sigma$ threshold at 95 and 150~GHz simultaneously, with 65 of those exceeding the 5$\sigma$ detection threshold at 95 and 150~GHz simultaneously. 

We can extrapolate this number to the SPT-3G Galactic Plane survey to estimate an expected event rate by making a few assumptions. Since all of the stellar flares we detected are local (within 1~kpc; see \autoref{fig:lum_compare}), we assume the local detectable stellar density is similar between the fields. Since the timescales of these flare events are typically longer than the cadence of observations but shorter than the survey durations, we assume the expected event rate scales with both the survey area and the survey duration. Under these assumptions, we would expect approximately 0.3 stellar flare events within the 500 hours of the SPT-3G Galactic Plane Survey data analyzed here.

The observed rate of two flares is higher than the expected rate of 0.3 flares. One possible explanation is that the SPT-3G Galactic Plane Survey can probe a larger volume, as it is not limited by the thickness of the Galactic Plane like the Main field survey, and can therefore detect bright sources at larger distances. The two detected sources indeed lie at larger distances and have higher luminosities compared to flares detected in the main field (\autoref{fig:lum_compare}). Given that the two bright flares would also be detectable in the main field, applying the same assumptions predicts hundreds of bright flares in the four-year main field survey, which we do not observe. This supports the hypothesis that the Galactic Plane Survey can find more bright sources by probing a larger volume and accessing new populations of sources. One caveat is that these estimates are based on small number statistics, as the sample from the SPT-3G Galactic Plane Survey is small, thus it is too early to draw any definitive conclusions from the comparison between this work and the main field transient search. Future data from the SPT-3G Galactic Plane Survey will hopefully provide a larger sample, improving our understanding of the mm-wave transient population in the Galactic Plane.

\section{Conclusion}
\label{Sec:conclusion}


This paper presents the first search for transients in the SPT-3G Galactic Plane Survey. In recent years, we have seen the emergence of a new subfield in mm-wave astronomy, namely the use of instruments designed to measure the CMB as blind mm-wave transient monitors. Previous efforts in this new discipline have all focused on the extragalactic sky; in this work we have presented a new survey targeting the Galactic Plane with the SPT-3G camera and we have reported the detection of two transient events at high significance. The SPT-3G Galactic Plane Survey covers approximately 100~$\text{deg}^2$, equally divided into three subfields, with Sgr A centered in the subfield at the lowest elevation. The observations analyzed in this paper were conducted between 2023 February 13 and 2023 March 12 and between 2024 March 1 and 2024 March 20. Typically, each subfield was observed about 13 times in each Galactic Plane observing day. In total, this analysis includes approximately 1,500 individual 20-minute observations with a total on-sky time of 500 hours. The median noise levels of the individual maps are 4.6, 6.6, and 24.6~mJy at 95, 150, and 220~GHz, respectively. The two-year map depths are 6.8, 8.3, and 20~$\mu\mathrm{K}$-arcmin (corresponding to 1.0, 1.6, and 4.0~mJy) at 95, 150, and 220~GHz, respectively.

A mapmaking pipeline similar to the one used in other analyses with SPT-3G is employed to generate a map for each observation, with modifications specifically tailored to the SPT-3G Galactic Plane Survey. A yearly average map is subtracted from each observation map in order to remove static backgrounds, including, but not limited to, the CMB, static sources, and emission from the Galaxy. Sources with a S/N greater than 5$\sigma$ in both 95 and 150~GHz bands simultaneously are identified using the transient detection pipeline. Lightcurves for the detected sources are extracted and analyzed, leading to the identification of two astrophysical transient sources, SPT-SV J174417.2-293942 (Source 1) and SPT-SV J173508.3-292956 (Source 2). The two transient events have the following characteristics:

(1) Both events have timescales of approximately one day. Source 1 has a timescale of 1.00~$\pm$~0.05 days. Source 2 has a timescale of 0.80~$\pm$~0.08 days.

(2) Source 1 has peak flux densities of 80.3~$\pm$~4.7~mJy at 95~GHz and 59.4~$\pm$~6.6~mJy at 150~GHz. Source 2 has peak flux densities of 58.7~$\pm$~4.5~mJy at 95~GHz and 86.6~$\pm$~6.2~mJy at 150~GHz.

(3) Neither source is detected in the yearly average maps. Each has a measured quiescent flux density consistent with local noise.

(4) No linear polarization has been detected for either event.

(5) Source 1 has an average spectral index of $\alpha^{95}_{150} = -0.29~\pm~0.04$, indicating a falling spectrum. Source 2 has an average spectral index of $\alpha^{95}_{150} = 0.92~\pm~0.03$, indicating a rising spectrum.

(6) We identify counterparts of both sources in other surveys, including \textit{Chandra}, \textit{Gaia}, 2MASS, and \textit{WISE}.

(7) Source 1 has a peak luminosity of $9.95~\times~10^{30}~\text{erg}~\text{s}^{-1}$ at 150~GHz. Source 2 has peak luminosity of $3.48~\times~10^{31}~\text{erg}~\text{s}^{-1}$ at 150~GHz.

Both sources are associated with binary systems containing white dwarf accretors---one with a subgiant companion and the other with a giant companion. The emission mechanism(s) for the flares in the millimeter wavelengths are unclear, though one plausible explanation could be self-absorbed synchrotron emission for both events, with the different spectral indices arising from differences in optical depth. The transient mm-wave emission likely originates from the accretion disks, though we do not yet have sufficient evidence to confirm this hypothesis. 

The 5$\sigma$ detection threshold used in this work is intentionally conservative due to the high noise fluctuations and contamination from bright sources in the Galactic field; however, this also limits our search to only bright transients. In the future, we plan to conduct a deeper transient search with the current and future data sets, lowering the detection threshold to 3$\sigma$ to capture fainter transients once the noise is better characterized. We also plan to expand the timescale range we are exploring. The method used in this work is sensitive to transients with timescales on the order of hours to days. We plan to extend this range on both ends---probing faster transients with timescales of minutes using single-scan maps, as well as longer transients with timescales of months to years using yearly maps. The SPT will continue to observe the Galactic Plane for approximately one month per year through the end of SPT-3G observations. Similar methods will be applied to future data and are expected to yield more transient detections. These observations have the potential to probe the rarest and most luminous Galactic transients and reveal differences in the density and luminosity distribution of transient sources when observed along lines perpendicular and parallel to the Galactic Plane.

\section{Acknowledgments}

TJM thanks Emily Leiner for useful discussions about the population size for RS CVn stars, and Margaret Ridder and Craig Heinke for useful discussions about cm-band flaring in cataclysmic variables.

The South Pole Telescope program is supported by the National Science Foundation (NSF) through awards OPP-1852617 and OPP-2332483. Partial support is also provided by the Kavli Institute of Cosmological Physics at the University of Chicago. 
Argonne National Laboratory’s work was supported by the U.S. Department of Energy, Office of High Energy Physics, under contract DE-AC02-06CH11357. 
The UC Davis group acknowledges support from Michael and Ester Vaida. 
Work at the Fermi National Accelerator Laboratory (Fermilab), a U.S. Department of Energy, Office of Science, Office of High Energy Physics HEP User Facility, is managed by Fermi Forward Discovery Group, LLC, acting under Contract No. 89243024CSC000002. 
The Melbourne authors acknowledge support from the Australian Research Council’s Discovery Project scheme (No. DP210102386). 
The Paris group has received funding from the European Research Council (ERC) under the European Union’s Horizon 2020 research and innovation program (grant agreement No 101001897), and funding from the Centre National d’Etudes Spatiales. 
The SLAC group is supported in part by the Department of Energy at SLAC National Accelerator Laboratory, under contract DE-AC02-76SF00515. 
This work was partially supported by the Center for AstroPhysical Surveys (CAPS) at the National Center for Supercomputing Applications (NCSA), University of Illinois Urbana-Champaign.

This research has made use of data obtained from the 3XMM XMM-Newton serendipitous source catalogue compiled by the 10 institutes of the XMM-Newton Survey Science Centre selected by ESA. This research has made use of data obtained from the Chandra Source Catalog, provided by the Chandra X-ray Center (CXC). 
This work has made use of data from the European Space Agency (ESA) mission
{\it Gaia} (\url{https://www.cosmos.esa.int/gaia}), processed by the {\it Gaia}
Data Processing and Analysis Consortium (DPAC,
\url{https://www.cosmos.esa.int/web/gaia/dpac/consortium}). Funding for the DPAC
has been provided by national institutions, in particular the institutions
participating in the {\it Gaia} Multilateral Agreement. 
This publication makes use of data products from the Two Micron All Sky Survey, which is a joint project of the University of Massachusetts and the Infrared Processing and Analysis Center/California Institute of Technology, funded by the National Aeronautics and Space Administration and the National Science Foundation. 
This publication makes use of data products from the Wide-field Infrared Survey Explorer, which is a joint project of the University of California, Los Angeles, and the Jet Propulsion Laboratory/California Institute of Technology, funded by the National Aeronautics and Space Administration. 

This scientific work uses data obtained from Inyarrimanha Ilgari Bundara, the CSIRO Murchison Radio-astronomy Observatory. We acknowledge the Wajarri Yamaji People as the Traditional Owners and native title holders of the Observatory site. CSIRO’s ASKAP radio telescope is part of the Australia Telescope National Facility (https://ror.org/05qajvd42). Operation of ASKAP is funded by the Australian Government with support from the National Collaborative Research Infrastructure Strategy. ASKAP uses the resources of the Pawsey Supercomputing Research Centre. Establishment of ASKAP, Inyarrimanha Ilgari Bundara, the CSIRO Murchison Radio-astronomy Observatory and the Pawsey Supercomputing Research Centre are initiatives of the Australian Government, with support from the Government of Western Australia and the Science and Industry Endowment Fund.
This paper includes archived data obtained through the CSIRO ASKAP Science Data Archive, CASDA (https://data.csiro.au/).

The Dark Energy Camera Plane Survey (DECaPS; NOAO Proposal ID 2016A-0323 and 2016B-0279, PI: D. Finkbeiner) includes data obtained at the Blanco telescope, Cerro Tololo Inter-American Observatory, National Optical Astronomy Observatory (NOAO).
The NSF NOIRLab is operated by the Association of Universities for Research in Astronomy (AURA) under a cooperative agreement with the National Science Foundation. Database access and other data services are provided by the ASTRO Data Lab.
This project used data obtained from the Dark Energy Camera (DECam), which was constructed by the Dark Energy (DES) collaboration. Funding for the DES Projects has been provided by the U.S. Department of Energy, the U.S. National Science Foundation, the Ministry of Science and Education of Spain, the Science and Technology Facilities Council of the United Kingdom, the Higher Education Funding Council for England, the National Center for Supercomputing Applications at the University of Illinois, Urbana-Champaign, the Kavli Institute of Cosmological Physics at the University of Chicago, the Center for Cosmology and Astro-Particle Physics at the Ohio State University, the Mitchell Institute for Fundamental Physics and Astronomy at Texas A\&M University, Financiadora de Estudos e Projetos, Funda\c{c}\~ao Carlos Chagas Filho de Amparo \`a Pesquisa do Estado do Rio de Janeiro, Conselho Nacional de Desenvolvimento Cient{\'i}fico e Tecnol{\'o}gico and the Minist{\'e}rio da Ci{\^e}ncia, Tecnologia e Inova\c{c}\~ao, the Deutsche Forschungsgemeinschaft, and the Collaborating Institutions in the Dark Energy Survey.
The Collaborating Institutions are Argonne National Laboratory, the University of California at Santa Cruz, the University of Cambridge, Centro de Investigaciones Energ{\'e}ticas, Medioambientales y Tecnol{\'o}gicas-Madrid, the University of Chicago, University College London, the DES-Brazil Consortium, the University of Edinburgh, the Eidgen{\"o}ssische Technische Hochschule (ETH) Z{\"u}rich, Fermi National Accelerator Laboratory, the University of Illinois, Urbana-Champaign, the Institut de Ci{\`e}ncies de l'Espai (IEEC/CSIC), the Institut de F{\'i}sica d'Altes Energies, Lawrence Berkeley National Laboratory, the Ludwig-Maximilians Universit{\"a}t M{\"u}nchen and the associated Excellence Cluster Universe, the University of Michigan, the National Optical Astronomy Observatory, the University of Nottingham, the Ohio State University, the OzDES Membership Consortium, the University of Pennsylvania, the University of Portsmouth, SLAC National Accelerator Laboratory, Stanford University, the University of Sussex, and Texas A\&M University.

This research was done using services provided by the OSG Consortium \citep{osg07, osg09}, which is supported by the National Science Foundation awards \#2030508 and \#2323298. This work relied on the \texttt{NumPy} library for numerical computations \citep{numpy}, the \texttt{SciPy} library for scientific computing \citep{scipy}, and the \texttt{Matplotlib} library for plotting \citep{matplotlib}.

\clearpage
\bibliography{sample631}{}
\bibliographystyle{aasjournal}



\end{document}